\documentclass[a4paper,11pt,openright]{thesis}  
\usepackage{superfig,morefloats}
\usepackage{graphicx}
\usepackage{amssymb}
\usepackage[dvips]{dropping}  
\usepackage{setspace}
\usepackage{section,relsize}
\usepackage{caption2,topcapt}
\usepackage[Lenny]{fncychap}  
\usepackage{xspace}           
\usepackage{xcolor}
\usepackage{sidecap}          
\usepackage{fancyhdr}
\usepackage{overrulehere}
\usepackage{natbib}
\usepackage[pdfpagelabels=false]{hyperref}	
\hypersetup{colorlinks=true, linkcolor=blue,
citecolor=blue,filecolor=blue,urlcolor=blue}
\usepackage{pdfpages}
\usepackage{titlesec}
\usepackage{upgreek}
\renewcommand{\captionlabeldelim}{---}
\renewcommand{\captionfont}{\footnotesize}
\renewcommand{\captionlabelfont}{\small \sc}
\newcounter{savesection}
  {\setcounter{savesection}{\value{section}}%
   \setcounter{section}{0}%
   \renewcommand\thesection{\thechapter.\Alph{section}}}%
  {\setcounter{section}{\value{savesection}}}
\captionstyle{normal}
\newcommand{\vsini}{\mbox{$v\sin i$}}
\newcommand{\vmac}{$v_{\rm mac}$}
\newcommand{\vmic}{$\xi$\xspace}

\newcommand{\Teff}{\mbox{$T_{\rm eff}$}}
\newcommand{\logg}{\mbox{$\log g$}}
\newcommand{\logQ}{\mbox{$\log Q$}}

\newcommand{\vcrit}{$v_{\rm crit}$}
\newcommand{\logLs}{$\log (\mathcal{L}/\mathcal{L}_{\odot})$}
\newcommand{\logL}{$\log (L/L_{\odot})$}
\newcommand{\Ls}{$ \mathcal{L}$}
\newcommand{\He}{$Y_{\rm He}$}

\newcommand{\difMs}{$M_{\rm sp}/M_{\rm ev(HR)}$}
\newcommand{\ls}{\mbox{$\lesssim$}\,}
\newcommand{\gs}{\mbox{$\gtrsim$}\,}

\newcommand{\kms}{\,\mbox{km\,s$^{-1}$}\xspace}
\newcommand{\MSol}{\,\mbox{M$_\odot$}\xspace}

\newcommand{\clearemptydoublepage}{\newpage{\pagestyle{empty}\cleardoublepage}}
\def\prefacesection#1{%
	\chapter*{#1}
	\addcontentsline{toc}{chapter}{#1}}
\begin{document}
\frontmatter
\includepdf[pages=-]{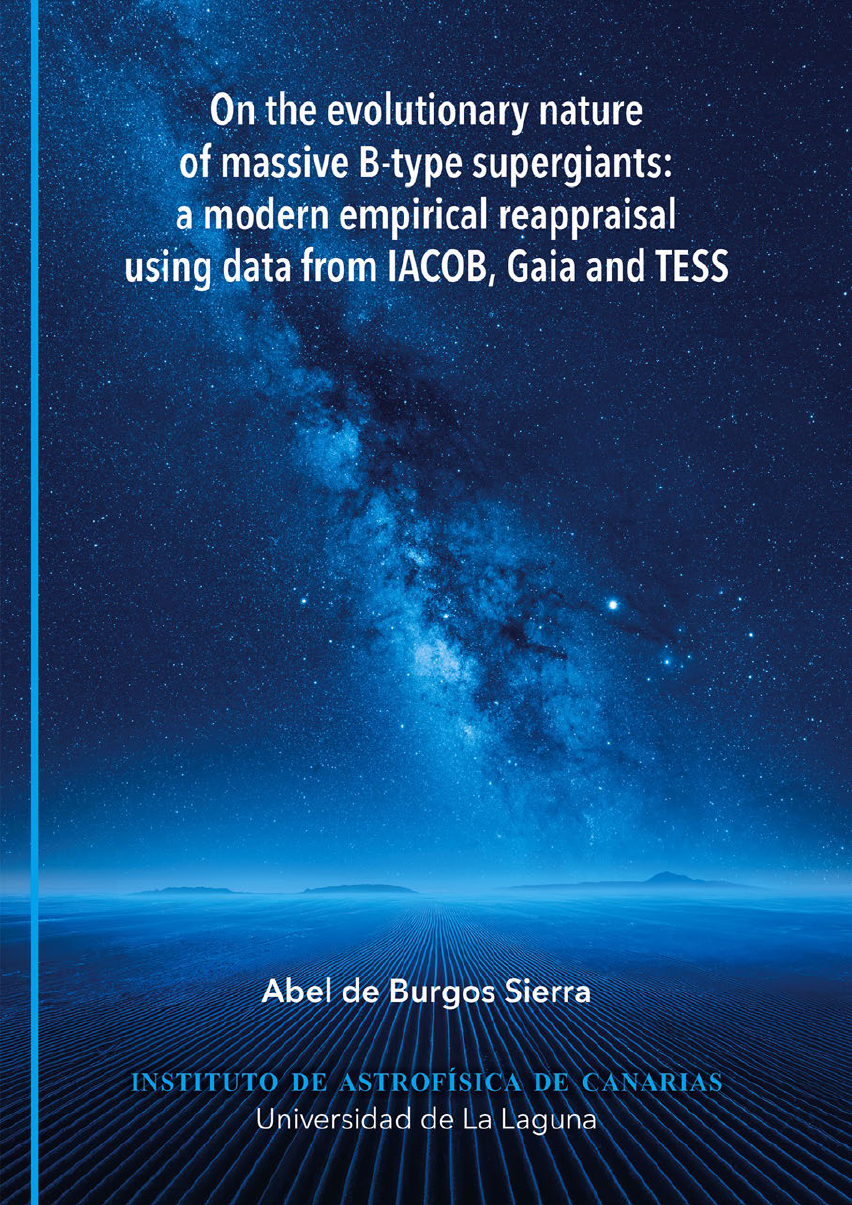}
\begin{minipage}[b]{0.45\linewidth}
{\hspace{-0.6cm} \large UNIVERSIDAD \de LA LAGUNA\\
\vspace{0.25cm}
\hspace{-0.72cm} Departamento de Astrofísica}
\end{minipage}
\hfill
\begin{minipage}[b]{0.39\linewidth} \includegraphics[width=6cm]{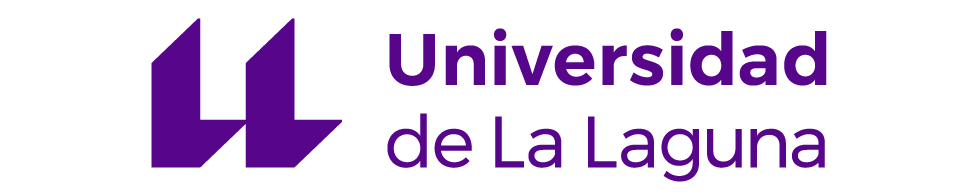}\vspace{0.1cm}
\end{minipage}
\begin{center}\rule{13cm}{0.4pt}\end{center}
%
%
\title{On the evolutionary nature of massive B-type supergiants: a modern empirical reappraisal using data from IACOB, \textit{Gaia} and \textit{TESS}}
%
%
\author{Abel de Burgos Sierra}
\titulobonito
\thispagestyle{empty}
\ \vfill\noindent
%
%
Examination date: October, 2024  \\
Thesis supervisors: Dr.~Sergio Simón-Díaz (IAC, University of La Laguna)\\
\textcolor{white}{Thesis supervisors:} Dr.~Miguel A. Urbaneja (University of Innsbruck)\\
 \\
\copyright{\,Abel de Burgos Sierra 2024}\\
%
%
%
%
%
\clearemptydoublepage{}
\thispagestyle{empty}
\begin{tabular}{cl}
 & 
\end{tabular}
\vspace{1cm}
%
%
{\flushright{\it To my family,\\
        for their infinite support.\/\\}}
\clearemptydoublepage{}
\thispagestyle{empty}
%
%
\clearemptydoublepage{}
%
%
\prefacesection{Resumen}
\vspace{-1.1cm}

Las estrellas masivas contribuyen decisivamente a la evolución química y dinámi-ca de las galaxias y el Universo. Su intensa radiación ionizante, fuertes vientos estelares y catastróficos finales como supernovas o estallidos de rayos gamma contribuyen a una importante retroalimentación del medio que les rodea. Son además útiles como indicadores de distancia y en estudios extragalácticos.

A pesar de su importancia, las comparaciones hechas entre los datos observacionales y los modelos teóricos de estrellas masivas han revelado varias discrepancias que desafían nuestra comprensión de estos objetos. Una de las principales incertidumbres tratadas en esta tesis es la sobredensidad de estrellas supergigantes de tipo B en el diagrama de Hertzsprung--Russell donde los modelos predicen el final de la fase de secuencia principal, o TAMS. Resulta incierto si es necesario redefinir la posición de dicho final, o si la sobredensidad es el resultado de poblaciones superpuestas que siguen caminos evolutivos diferentes.

Inicialmente concebidas como descendientes directas de las estrellas de tipo O, las supergigantes B pueden incluir objectos que no sólo evolucionan en la secuencia principal, sino que también estén regresando de una fase post-supergigante roja. A su vez, la alta fracción de estrellas masivas que se espera que nazcan en sistemas binarios o múltiples crea canales adicionales en los que una fracción representativa de estas estrellas pueden ser el producto de interacciones binarias.
Además, algunas propiedades fundamentales de las supergigantes de tipo B no están tan bien definidas como en el caso de las estrellas de tipo O. 
Entre ellas se incluyen las propiedades rotacionales y la tasa de pérdida de masa, las cuales tienen un impacto significativo en la evolución de las estrellas masivas, así como la fracción de sistemas binarios.

Para superar esta situación y comparar adecuadamente los resultados empíri-cos con las predicciones teóricas se necesitan muestras estadísticamente representativas. En este sentido, los grandes conjuntos de datos espectroscópicos ofrecen una oportunidad única para estudiar las propiedades físicas y químicas de las supergigantes B, objetivo principal de esta tesis.
Además, la llegada de misiones espaciales de astrometría y fotometría como \textit{Gaia} y TESS marca una nueva era para estudiar propiedades adicionales con un detalle sin precedente.

Esta tesis comprende el estudio de casi un millar de supergigantes azules galácticas (de tipo O y B) combinando datos espectroscópicos multi-época de alta resolución del proyecto IACOB y del archivo de la ESO, con distancias de \textit{Gaia} y fotometría de TESS, convirtiéndose así en el mayor estudio empírico y holístico de las propiedades físico-químicas y pulsacionales de estos objetos realizado hasta la fecha.
Para alcanzar este objetivo, una parte significativa pero crucial de este trabajo se ha dedicado a lograr un nivel alto de homogeneidad y completitud en la muestra, la cual inicialmente comprendía 4000 espectros ópticos de unas 400 estrellas de tipo O9\,--\,B9, y finalmente alcanzó 9000 espectros de 980 de estos objetos. Esto se logró mediante la revisión individual de cada espectro utilizando el programa \textit{pyIACOB} desarrollado durante la tesis y la exitosa ejecución de varias campañas observacionales. 

El análisis espectroscópico cuantitativo de los datos se ha realizado utilizando dos herramientas semiautomáticas diferentes. El análisis de ensanchamiento de las líneas, utilizado para obtener la velocidad rotacional proyectada y la \textit{macroturbulencia}, se logró empleando la herramienta {\tt iacob-broad}. Los parámetros restantes, a saber, $T_{\rm eff}$, $\log g$, $\xi$, $\log Q$, y las abundancias químicas superficiales de helio, silicio, carbono, nitrógeno y oxígeno, se obtuvieron utilizando un emulador estadístico para espectros sintéticos de \textsc{FASTWIND}, combinado con un método Monte Carlo basado en cadenas de Markov. Los datos multi-época también permitieron identificar los systemas binarios espectroscópicos de una y dos líneas dentro de la muestra.

Todos estos parámetros combinados en una muestra única de volumen limitado permitieron proporcionar una reevaluación empírica de las principales propiedades de las supergigantes B. Además, se abordaron dos cuestiones fundamentales: la localización de la TAMS con respecto a los modelos evolutivos más utilizados, y la existencia del teóricamente predicho y ampliamente aceptado aumento de las tasas de pérdida de masa en la región de bi-estabilidad del viento, ambas con importantes implicaciones en la evolución de las estrellas masivas.
En el penúltimo capítulo, se investiga si toda esta información puede proporcionar nuevas pistas para desentrañar la intrincada naturaleza de las supergigantes B, para lo que también se incluye un trabajo piloto sobre las propiedades pulsacionales de la muestra.

Por último, pero no por ello menos importante, los conocimientos adquiridos durante esta tesis sin duda nos ayudarán a emprender futuros estudios dedicados a propiedades o submuestras específicas. En combinación con los datos de los próximos grandes sondeos espectroscópicos, es de esperar que lo aprendido aquí nos lleve a mejorar aún más nuestra comprensión de las propiedades fundamentales y la naturaleza de las estrellas masivas.

\clearpage
%
%
\prefacesection{Abstract}
\vspace{-0.7cm}

Massive stars are key contributors to the chemodynamical evolution of galaxies and the Universe. Their intense ionizing radiation, strong stellar winds, and extreme final fates, including core-collapse supernova or gamma-ray bursts, all contribute to an important feedback to their surrounding space. Furthermore, they represent useful tools as distance indicators and for extragalactic studies.

Despite their significance, comparisons between observational data and theoretical models of massive stars have revealed long-standing and new discrepancies that challenge our understanding of these objects. One major uncertainty addressed in this thesis is the overdensity of B-type supergiants in the Hertzsprung--Russell diagram, where models predict the termination of the main sequence phase, or TAMS. It became uncertain whether the location of the TAMS needs to be redefined or if the overdensity results from overlapping populations following different evolutionary paths.

Initially conceived as direct evolutionary descendants of O-type stars, B-type supergiants may include stars not only evolving in the main sequence but also returning from a post-red supergiant phase. Furthermore, the large fraction of massive stars expected to be born in binary or multiple systems created additional channels in which a representative fraction of these stars are predicted to be the products of binary interaction.
In addition, some fundamental properties of B-type supergiants are not as well constrained as in the O-type stars domain. These involve the spin-rate properties and the mass-loss rates, both of which have a significant impact on massive star evolution, as well as their multiplicity or binary fraction. 

To overcome this situation and effectively compare empirical results with different theoretical predictions, statistically significant samples are required. In this regard, large spectroscopic datasets offer a unique opportunity to study both the physical and chemical properties of B-type supergiants, which is the main purpose of this thesis.
Moreover, the advent of space astrometry and photometry missions such as \textit{Gaia} and TESS, has brought a new era for studying additional properties in unprecedented detail.

This thesis has comprised the study of almost a thousand Galactic blue supergiants (O- and B-type) combining multi-epoch high-resolution spectroscopic data from the IACOB project and the ESO archive with \textit{Gaia} distances and TESS photometry, becoming the largest holistic empirical study of the physical, chemical, and pulsational properties of these objects performed to date.
To reach this, a significant but crucial part of this work has been devoted to achieve a high level of homogeneity and completeness in the sample, which initially comprised 4000 optical spectra of about 400 O9\,--\,B9-type stars, and finally reached 9000 spectra of 980 of these objects. 
This was done by individual revision of the spectra using \textit{pyIACOB} own developed program and the successful execution of several observational campaigns. 

The quantitative spectroscopic analysis of the data has been performed using two different semi-automated tools. The line-broadening analysis, used to derive the projected rotational velocity and the \textit{macroturbulence}, was achieved using the {\tt iacob-broad} tool. The remaining parameters, namely $T_{\rm eff}$, $\log g$, $\xi$, $\log Q$ and the surface abundances of helium, silicon, carbon, nitrogen, and oxygen, were obtained using a statistical emulator for \textsc{FASTWIND} synthetic spectra, combined with a Markov chain Monte Carlo method. Additionally, the multi-epoch data let for the identification of single- and double-line spectroscopic binaries in the sample.

All these parameters gathered into a unique volume-limited sample allowed to provide an empirical reassessment of the main properties of B-type supergiants. Furthermore, two fundamental questions were addressed: the location of the TAMS with respect to the most-used evolutionary models, and the existence of the theoretically predicted and widely accepted increase in mass-loss rates over the wind bi-stability region, both of which have important implications in the evolution of massive stars.
In the penultimate chapter, it is investigated whether all this information can provide new clues to disentangle the intricate nature of B-type supergiants, for which a pilot work on the pulsational properties of the sample is also included.

Last but not least, the knowledge gained during this thesis will undoubtedly contribute to undertake future dedicated studies on specific properties or sub-samples. Combined with the data of upcoming large spectroscopic surveys, what we have learned here will hopefully lead us to further improve our understanding of the fundamental properties and nature of massive stars.
\clearpage
\prefacesection{Agradecimientos}
\vspace{-0.5cm}

Mi más sentido agradecimiento es para mi abuela Carmen y mi abuelo Mariano, quienes hicieron posible que, con apenas 6 años, ya apuntase mi primer telescopio hacia las estrellas. Gracias abuela por llevarme a ver aquel cielo de verano en Ariza, ese que me cautivó. Quién nos lo iba a decir...

Gracias de todo corazón a mis padres por su infinito apoyo, su infinita paciencia, y su infinito amor. Por además mantener la llama de la astronomía viva durante todos aquellos años, especialmente gracias a todos los viajes que hicimos con el coche y el equipo para observar o fotografiar el cielo. 

Al resto de mi familia también, pues aunque he estado lejos estos años, sé que me apoyáis en todo, y deseáis lo mejor para mí.

Por supuesto, al responsable de que esta tesis haya sido posible, el Dr. Sergio Simón-Díaz. Hace justo seis años que nos conocemos, y nunca me ha faltado ni tu paciencia ni tu apoyo. Has cuidado de mí en todos los sentidos, y me siento infinitamente agradecido por ello. Gracias, además, por haberme abierto tantas puertas, por haberme mantenido motivado en todo momento, por haberme enseñado tanto. Ojalá algún día volvamos a observar juntos en La Palma. Gracias igualmente a mi codirector, el Dr. Miguel A. Urbaneja. Esta tesis no hubiese sido posible de no ser por tu código MAUI. Además de apoyarme siempre, me has enseñado a ser más crítico y objetivo con mis resultados, y, en definitiva, a mejorar la calidad científica de toda esta tesis y de mi trabajo futuro. También te estoy muy agradecido por tu hospitalidad durante las estancias en Innsbruck. Espero que tengas alguna ruta en mente para cuando vuelva a visitarte.

Quiero extender mi agradecimiento a muchas personas que me acompañaron en estos seis años en Canarias, dos en La Palma, cuatro en Tenerife. Esta experiencia no hubiera sido igual sin ellos. Gracias a Cecilia Fariña por inspirarme y ser mi referente en los observatorios. A todos los estudiantes del grupo NOTING por todos los planes y vivencias en La Palma. Lo que daría por una última fiesta en la terraza de El Drago junto con Luke. A mi amigo Vladimir, con quien he compartido importantes momentos de mi vida, también en La Palma. A mi amiga Leona, por todas esas excursiones alrededor de nuestra querida Isla Bonita. También a Lucía, de quien aprendí tantísimas cosas de la vida en esos años. Ya en Tenerife, quiero agradecer especialmente a Ana, Lucía, y Sandra por acogerme en el grupo de intrépidas excursionistas, al menos 2/6 de ellas. Junto con Alejandra, Andreu, Jaume, Jorge y Josu, y más adelante Ester, Pablo y Yessica, formaron la que para mí ha sido mi familia durante esta tesis. Os tengo infinita gratitud y no sabéis lo afortunado que me siento por todos los momentos, rutas, y viajes compartidos. Por supuesto, no puedo dejar pasar por alto la 040 junto con Lucía y Jaume, hito que no hubiese sido posible de no ser por esta gran familia que nos apoyó en todo momento. Tampoco el apoyo de Pablo durante esta última etapa. Gracias por tu paciencia en el ``gastropiso" y por preocuparte por mi. Gracias a mis compañeros y compañeras de oficina - Mar, Martín, Natalia, Zahra... - por vuestro apoyo; igualmente a Bea, Emma, Mónica, Nacho, Teresa, Z$\upomega\acute{\upeta}$; a todo el equipo de la IACOB; a mis recientes amistades latinoamericanas con Ángela, Enzo y Matías; y por último, a todas aquellas personas que saben que deberían estar aquí pero, por falta de memoria o tiempo, no han aparecido, por favor, perdonadme.

Agradecer a mi otra familia, la de expertos en estrellas masivas, que durante estos años no han hecho sino inspirarme a mejorar y aprender de este bonito campo de la astrofísica. Especialmente a Artemio Herrero, Francisco Najarro, Ignacio Negueruela, Joachim Puls, Rolf Kudritzki, Norbert Przybilla, Sara Rodríguez y Zsolt Keszthelyi, pero también al resto y a los estudiantes.

Quiero agradecer a Juan Carlos y a Montse por todos los valiosos consejos que me habéis dado, no solo los de viajes. Por vuestra hospitalidad, y vuestro buen trato donde fuese que nos encontrásemos. Siempre deseando escucharos.

Gracias en forma de risas a Andreu Buenafuente y Berto Romero. Vuestro programa ``Nadie Sabe Nada" ha sido mi válvula de escape en muchos fines de semana. Esta tesis tiene bastante ``samanté". 

A K$\upalpha\uptau\upepsilon\uprho\acute{\upiota}\upnu\upalpha$, por haber cuidado tanto de mí y haberme dado todos los ánimos que necesitaba para seguir en los momentos más difíciles. Has sido uno de mis grandes descubrimientos durante esta tesis. E$\upsilon\upchi\upalpha\uprho\upsilon\upiota\upsigma\uptau\acute{\upomega}$ 
$\upkappa\upalpha\uprho\updelta\upiota\acute{\alpha}$ $\upmu$o$\upsilon$.

Mi agradecimiento al NOT y al IAC, cuyas becas me han permitido llevar a cabo este trabajo y conocer a tantísimos grandes profesionales. También a Artemio Herrero, Dominic Bowman y Norbert Przybilla por sus valiosos comentarios a la tesis.

%
%
{\flushleft{Abel de Burgos Sierra}}

\clearpage
\prefacesection{Acknowledgments}
\vspace{-0.5cm}

My most profound gratitude goes to my grandmother Carmen and my grandfather Mariano, who made it possible for me to point my first telescope at the stars when I was only 6 years old. Thank you grandma for taking me to see the summer sky in Ariza, the one that captivated me. Who was going to tell us...

A heartfelt thanks to my parents for their infinite support, their infinite patience, and their infinite love. For also keeping the flame of astronomy alive during all those years, especially thanks to all the trips we made with the car and the equipment to observe or photograph the sky. 

To the rest of my family as well, because even though I have been far away these years, I know that you always support me, and wish the best for me.

Of course, to the person responsible for making this thesis possible, Dr. Sergio Simón-Díaz. We have known each other for exactly six years and I have never lacked your patience or your support. You have taken care of me in every way, and I feel infinitely grateful for that. Thank you also for having opened so many doors for me, for having kept me motivated at all times, for having taught me so much. I hope that someday we will observe together again in La Palma. Thanks also to my co-supervisor, Dr. Miguel A. Urbaneja. This thesis would not have been possible without your MAUI code. In addition to always supporting me, you have taught me to be more critical and objective with my results and, ultimately, to improve the scientific quality of this thesis and my future work. I am also very grateful for your hospitality during the stays in Innsbruck. I hope you have some route in mind for when I visit you again.

I want to extend my thanks to many people who accompanied me in these six years in the Canaries, two in La Palma, four in Tenerife. This experience would not have been the same without them. Thanks to Cecilia Fariña for inspiring me and being my reference at the observatories. To all the students of the NOTING group for all the plans and experiences in La Palma. I wish there was one last party on the terrace of El Drago together with Luke. To my friend Vladimir, with whom I have shared important moments of my life, also in La Palma. To my friend Leona, for all those excursions around our beloved Isla Bonita. Also to Lucia, from whom I learned so many things about life during those years.
Once in Tenerife, I want to especially thank Ana, Lucia, and Sandra for accepting me in your group of intrepid hikers, at least 2/6 of you were. Together with Alejandra, Andreu, Jaume, Jorge and Josu, and later Ester, Pablo and Yessica, they formed what for me has been my family during the thesis. I have infinite gratitude, and you do not know how fortunate I feel for all the moments, routes, and trips we shared. Of course, I cannot skip the 040 with Lucía and Jaume, a milestone that would not have been possible without that great family that supported us all the way. Thank you Pablo for your support during this last stage of the thesis, for your patience in our ``gastropiso" and for caring about me. Thanks to my office mates - Mar, Martín, Natalia, Zahra... - for your support; also to Bea, Emma, Mónica, Nacho, Teresa, Z$\upomega\acute{\upeta}$; to all the IACOB team; to my recent Latin American friendships Ángela, Enzo and Matías; and finally, to all those people who know they should be here but, due to my lack of memory or time, have not appeared, please forgive me.

I would like to thank my other family, the experts in massive stars, who during these years have done nothing but inspire me to improve and learn from this beautiful field of astrophysics. Especially to Artemio Herrero, Francisco Najarro, Ignacio Negueruela, Joachim Puls, Rolf Kudritzki, Norbert Przybilla, Sara Rodríguez and Zsolt Keszthelyi, but also to the rest and to the students.

I want to thank Juan Carlos and Montse for all the valuable advice you have given me, not only about traveling. For your hospitality and your care wherever we met. Always looking forward to hearing from you.

Thanks in the form of laughs to Andreu Buenafuente and Berto Romero. Your program ``Nadie Sabe Nada" has been my escape valve on many weekends. This thesis has a lot of ``samanté".

To K$\upalpha\uptau\upepsilon\uprho\acute{\upiota}\upnu\upalpha$, for taking such good care of me and giving me all the encouragement I needed to continue in the most difficult moments. You have been one of my greatest discoveries during this thesis. E$\upsilon\upchi\upalpha\uprho\upsilon\upiota\upsigma\uptau\acute{\upomega}$ 
$\upkappa\upalpha\uprho\updelta\upiota\acute{\alpha}$ $\upmu$o$\upsilon$.

My thanks to the NOT and the IAC, whose grants have allowed me to carry out this work and to meet so many great professionals. Also to Artemio Herrero, Dominic Bowman and Norbert Przybilla for their valuable comments on the thesis.

%
%
{\flushleft{Abel de Burgos Sierra}}

\clearemptydoublepage{}
%
%
\tableofcontents
\clearemptydoublepage{}
\mainmatter{}
%
%
\large
\clearemptydoublepage{}

%
%
\chapter{Introduction}
%
%
\label{intro}
{\flushright{\it Science is more than a body of knowledge,\\
                  it is a way of thinking.\\\smallskip}
             {\rm \small Carl Sagan\\}}

\vspace{0.7cm}

\section{Massive stars}
\label{intro.1.massivestars}

\subsection{The importance of massive stars in the Universe}
\label{intro.1.1.importance}

\begin{figure}
    \centering
    \includegraphics[width=0.9\linewidth]{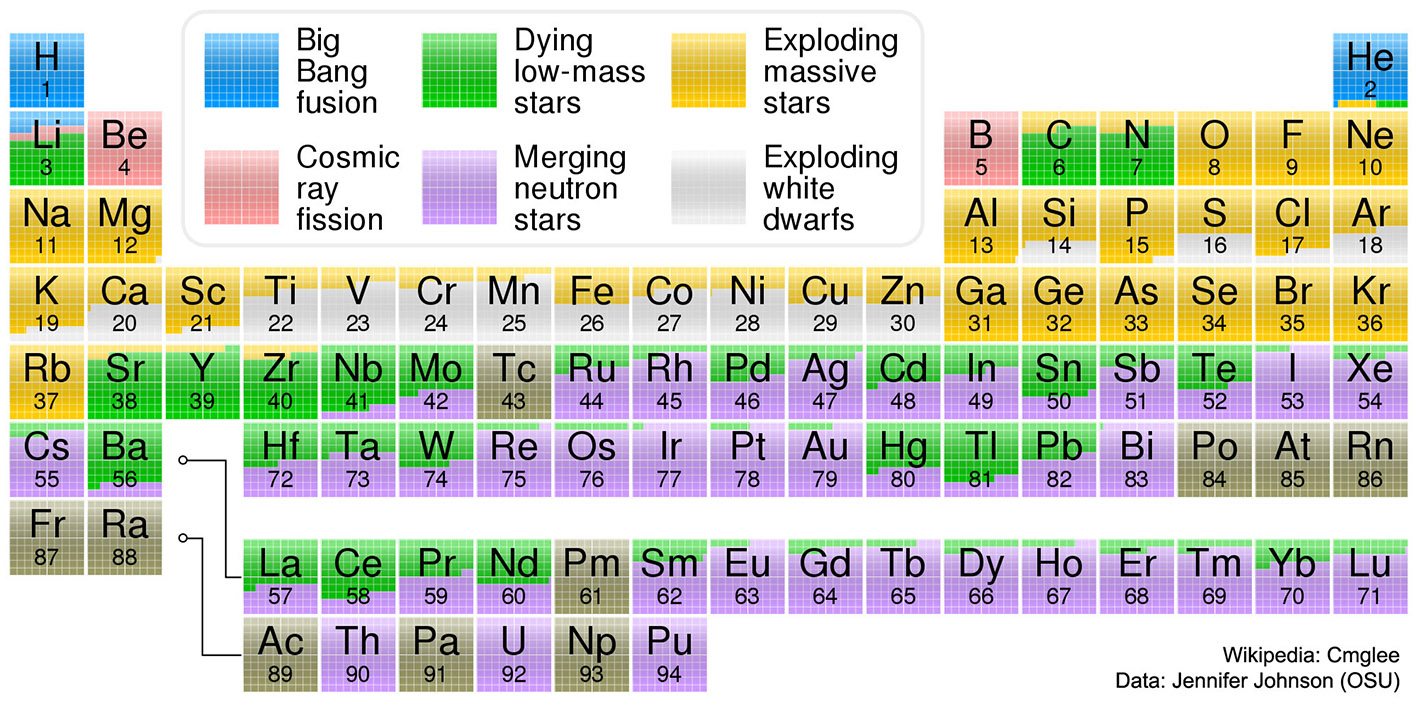}
    \caption{Periodic table displaying the currently believed origins of each atomic element. Stars in a wide range of masses are also able to produce elements from carbon to sulfur by charged-particle fusion reactions. The iron group elements are likely formed during the event of supernovae explosions. Further elements would be produced by neutron capture in massive stars by different proposed mechanisms still under debate. Credit: CMG Lee based on the original image by Jennifer Johnson (OSU).}
    \label{fig:elem}
\end{figure}

\dropping[0pt]{2}{M}{\sc assive} stars play a pivotal role in shaping the Universe as we know it. Born with initial masses of eight times that of the Sun or more, they are the cosmic engines that drive the evolution of galaxies and the production of heavy elements \citep{matteucci08, hopkins14, eldridge22}.
During the formation stages of their lives, they become the dominant source of energy, which is liberated into their parental molecular clouds, leading to the formation, but also to the destruction of their still forming siblings \citep[e.g.,][]{hollenbach94, scally01, matzner02}.
Compared to less-massive stars, their convective cores are able to fuse heavier elements, producing elements such as oxygen, sodium, phosphorus, sulfur, or iron, all of which are essential for the formation of planets and life as we know it \citep[e.g.,][see also Fig.~\ref{fig:elem}]{woosley15, hirschi15, kobayashi20}.
Their higher core mass, temperature, and pressure are also key aspects during the final stages of their evolution, leading to the production of some of the most extreme cosmic events, such as core-collapse supernovae and gamma-ray bursts \citep[e.g.,][]{woosley06, smartt09}, which further extend the production of heavy elements. While the remnants of these catastrophic events - neutron stars and black holes - represent some of the most exotic objects in the Universe, the ejected material becomes an important source of chemical enrichment in galaxies \citep[e.g.,][]{matteucci86, nomoto13, nomoto17}. Furthermore, the release of large amounts of gas and debris is crucial in the formation of the next generation of stars and planets \citep[e.g.,][]{herbst77, walter94, ansdell20}. 

The enormous amount of energy produced in their cores makes massive stars among the hottest and most luminous objects in the Universe, with surface temperatures that can reach several tens of thousands of Kelvin and luminosities that can exceed a million times that of the Sun. These properties make them suitable for extragalactic studies, providing additional information on star formation and chemical abundances beyond the Milky Way \citep[e.g.,][]{monteverde00, urbaneja05a, urbaneja05b, bresolin16}. Their luminosities are so high that they can be used as distance indicators up to several megaparsecs \citep[][]{kudritzki00, kudritzki03b, kudritzki08}, as well as to study very low-metallicity environments \citep[e.g.,][]{evans07, vink23}, some even close to the conditions of the early Universe \citep{Hirschi07, garcia19}. Furthermore, they were fundamental for the formation of the first galaxies \citep{Bromm04, Robertson10}, playing an important role in the epoch of reionization of the Universe \citep[see][]{Bromm01, Heckman11}.

Another important feature throughout the lives of massive stars is the presence of intense stellar winds. These can reach supersonic velocities of up to 2000\,--\,3000\,km\,s$^{-1}$ and expel a significant amount of their initial mass at rates of up to $10^{-5}$\,--\,$10^{-4}$\,M$_{\odot}$\,yr$^{-1}$ \citep[see][and references therein]{vink22}, which becomes an additional source of chemical enrichment \citep[e.g.,][]{maeder81, woosley95, Cescutti10}. Combined with intense ultraviolet radiation and episodic eruptions, these winds play a crucial role in shaping the surrounding environment, as they are able to trigger or cancel out new episodes of star formation \citep[e.g.,][]{krause13, watkins19, kim19, geen21}.

Although the influence of massive stars on the interstellar medium and the chemical evolution of their host galaxies is undeniable, these stars are not the most common ones in the Universe. In fact, they only represent a small fraction of the total stellar population \citep{salpeter55, chabrier05}. In the Milky Way, for example, they represent less than 1\% of the total number of stars \citep{kroupa01}. Furthermore, their lifetimes span only a few million years, which is significantly shorter compared to their less-massive counterparts that can live for billions of years. Despite these facts, they are responsible for the majority of the energy output of galaxies \citep[e.g.,][]{leitherer99, shapley03, marques-chaves20}.

It is now well known that a significant fraction of massive stars interact with a companion at some point during their lifetimes \citep[][]{sana12, sana13, moe17}. These systems are of particular interest because they can lead to a variety of phenomena such as mass transfer, contact binary evolution, and stellar mergers that might represent important evolutionary channels to better understand the evolution of massive stars \citep{marchant23}. Moreover, the merging of black-hole or neutron-star systems has recently gained much attention in the astrophysics community because of associated production of gravitational waves \citep{abbott16, belczynski16, marchant16}.

\subsection{The formation and birth of massive stars}
\label{intro.1.2.formation}

Massive stars are typically found in young clusters and associations distributed in their host galaxies. In particular, in those where the fraction of gas is higher \citep[such as spiral or irregular galaxies; see][]{lada03, deWit04, deWit05}. This leads to the basic question of how massive stars form, which is crucial for understanding their evolution and final fates. 
Despite the fact that our understanding of how they form is not yet complete, the initial steps likely follow a pathway similar to their intermediate- and low-mass counterparts, although on much shorter timescales \citep{Davies11, mottram11}. This first involves the gravitational collapse of dense (10$^{4}$\,--\,10$^{5}$ particles/cm$^{-3}$) and cold (~10\,--\,20\,K) giant molecular clouds (GMCs) made mainly of molecular hydrogen, which can hold 10$^{4}$\,--\,10$^{6}$\,M$_{\odot}$ \citep[see reviews by][]{krumholz19, klessen23}. The next step in the formation of stars is the fragmentation of the GMCs, with sizes of 10\,--\,100\,pc, into smaller discrete clumps in a hierarchical process \citep{blitz86}. These clumps continue to collapse under their own gravity as long as their mass exceeds the associated Jeans mass of the region, reaching masses of 10\,--\,1000\,M$_{\odot}$ and sizes of 0.1\,--\,1\,pc \citep{murray10}. At this point, the gas of the clumps becomes more optically thick and less efficient in radiating away the energy gained by the release of gravitational potential energy, which leads to an increase in temperature (to 60\,--\,100\,K) and density (10$^{-13}$\,g\,cm$^{-3}$). While the gas at the core of the clumps becomes hotter, the outward pressure increases until the further gravitational collapse is halted. The core reaches a hydrostatic equilibrium, and the clumps then become protostars, which can still accrete mass from the surrounding material. Eventually, the temperature at the core of the protostar will reach 15 million Kelvin, and nuclear fusion of hydrogen into helium will start. The newly formed star has reached the zero-age main sequence (ZAMS), and its internal structure is composed of a convective core surrounded by a radiative envelope. It is important to note that, despite the very large mass of the initial GMCs, the efficiency of star formation is low, with only 0.2\% to 20\% of the total mass being converted into stars \citep{evans91, evans09, murray10}.

Beyond this simplified version of star formation, the process is more complex and involves a variety of other physical processes, such as turbulence, metallicity, magnetic fields, and feedback from the newly formed stars, which can have a significant impact on the properties of stars reaching the main sequence \citep[e.g.,][]{Orkisz17, Pattle23}. Moreover, the initial collapse of the GMCs can be triggered by a variety of mechanisms, such as the compression of the gas by supernova explosions, the collision of clouds, or the interaction with other galaxies.

However, even the simplified formation of massive stars is not yet fully understood. One of the main challenges is the fact that their formation requires a significantly larger accretion mass compared to less-massive objects. One possibility is that high-mass protostars keep accreting enough mass even when the nuclear fusion in their cores has already started \citep{Haemmerle19}, or they do so at high accretion rates \citep{Hosokawa09}. However, their stronger initial feedback favors the dispersion of the surrounding material on the same timescale as the star forms, limiting its final mass \citep[e.g.,][]{Vazquez-Semadeni19}. Other possibilities include episodic mass accretion \citep[e.g.,][]{hunter17, Elbakyan21}, or the merging with other nearby forming stars \citep[e.g.,][]{smith12, Stacy13}.
This situation is further complicated by the fact that massive stars at the ZAMS are still highly embedded in their parental molecular clouds \citep{zinnecker07, langer12}, making them much less accessible before they become visible at optical and ultraviolet wavelengths \citep{bernasconi96}. Therefore, our interpretation mainly depends on the observed properties when massive stars have already evolved on the main sequence \citep[][see below]{yorke86}.

\vspace{-0.15cm}
\subsection{The main sequence phase}
\label{intro.1.3.ms}

The main sequence phase of massive stars is the most stable phase of their evolution \citep{maeder00, heger00}. During this phase, the star is able to fuse hydrogen into helium in its core, producing an outward thermal pressure that counteracts the inward pressure of gravitational collapse from the overlying layers. It is in this phase where they spend most of their lives, which are significantly shorter compared to stars of lower masses. As an example, for initial masses around 120M$_{\odot}$, this phase only lasts $\sim$3 million years, whereas for a 12M$_{\odot}$ star, it extends to $\sim$20 million years \citep[see][]{ekstrom12}. In contrast, the Sun, with 1M$_{\odot}$, will spend $\sim$10 billion years on the main sequence. The reason for the much shorter timescales of massive stars is connected with the carbon–nitrogen–oxygen (CNO) cycle. This cycle is more efficient and has a higher energy production rate than the proton–proton chain, which is the dominant mechanism in less massive stars. The energy produced in the convective core is then transported to the radiative envelope and to the surface of the star, where it is radiated away. 

\begin{figure}[t!]
    \centering
    \includegraphics[width=0.9\linewidth]{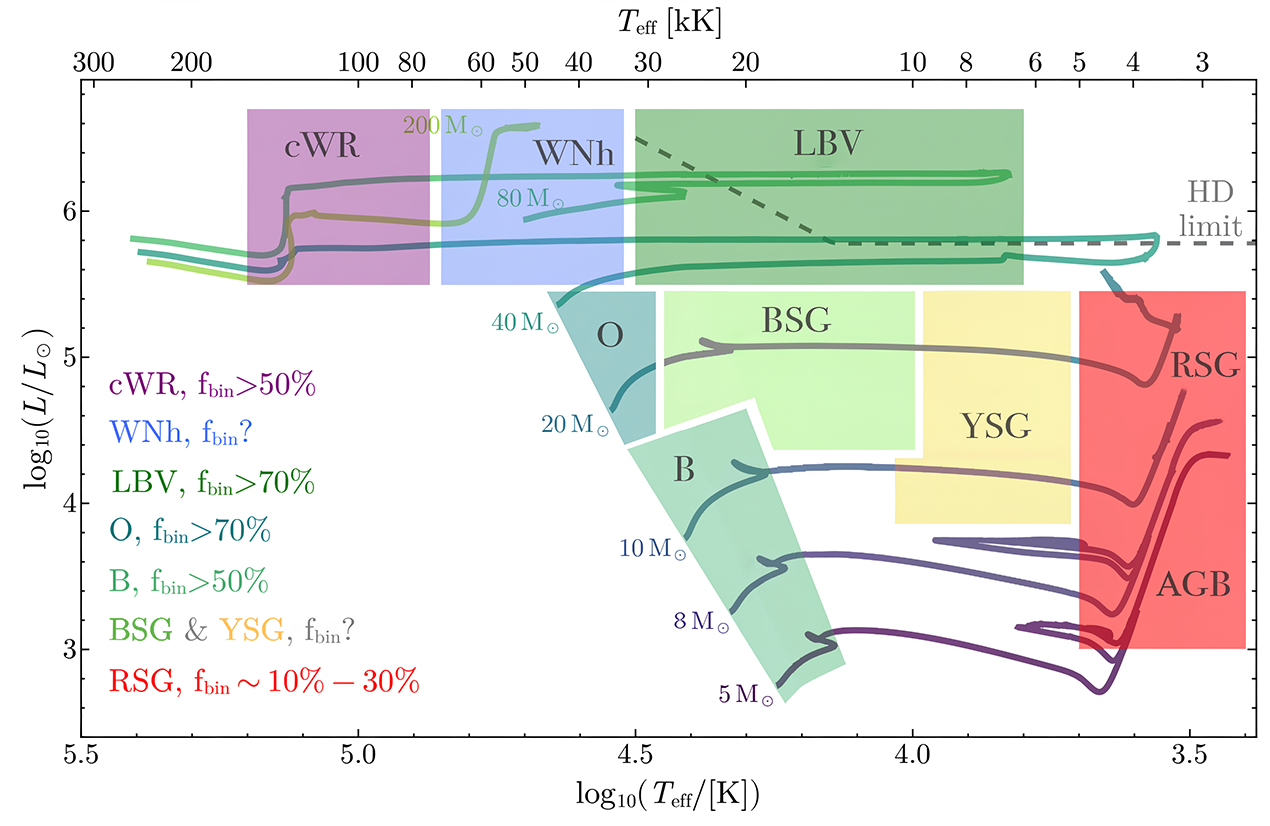}
    \caption{Schematic distribution of different evolutionary phases of massive stars in the Hertzsprung--Russell diagram. The known binary fraction is also included. The Humphreys-Davidson limit is indicated with dashed line \citep{humphreys79}. The AGB phase corresponding to stars of lower masses is combined with the RSG phase. Adapted figure from the review by \citet[][Fig.~1; reproduced with permission]{marchant23}. \vspace{-0.2cm}}
    \label{fig:HR_Marchant}
\end{figure}

Massive stars typically enter the main sequence phase as O- (M\,\gs18M$_{\odot}$) or early B-type dwarfs (8\,M$_{\odot}$ \ls M \ls 18\,M$_{\odot}$). Their effective temperature ranges between 45\,000\,K and 25\,000\,K, and their luminosities are between 10$^{4}$ and 10$^{6}$ times that of the Sun (see Fig.~\ref{fig:HR_Marchant}). During the evolution on the main sequence, the core undergoes a contraction and a temperature increase. At the same time, the higher energy rates produce the expansion of the outer layers of the star, leading to a decrease of the surface temperature and an increase in luminosity. The decrease in surface temperature along the main sequence is more significant for O-type stars than for the early B-type stars. 
Massive stars with 8\,M$_{\odot}<$ M $<$ 60\,M$_{\odot}$ evolve toward the end of the main sequence at a relatively constant luminosity. This phase ends when the hydrogen in the core is exhausted. At this point, different scenarios are proposed for the evolution of these objects that depend on the range of masses considered \citep[see][]{conti75, bond84, langer94, crowther95}. For stars up to $\sim$25\,M$_{\odot}$, they are predicted to evolve into red supergiants (RSGs), which is further supported by empirical observations of these objects up to those masses \citep[see][]{levesque05}. For stars with 25\,M$_{\odot}<$ M $<$ 60\,M$_{\odot}$, the situation is much less constrained. It has been proposed that these objects undergo a very short RSG phase, or simply skip it, and continue to evolve into luminous blue variables \citep[LBVs;][]{grafener12, groh13, weis20} and Wolf-Rayet \citep[WR;][]{conti75, abbott87, higgins21} stars\footnote{WNh in Fig.~\ref{fig:HR_Marchant} refers to WR stars showing hydrogen lines in their spectra, and are proposed to be progenitors of LBV stars \citep[see][]{conti75}.}. For stars more massive than $\sim$60\,M$_{\odot}$, the strong stellar winds and associated mass loss combined with instabilities caused by proximity to the Eddington limit \citep{sanyal15} make the evolution of these objects highly complex and uncertain.
Despite this simplified scheme, it is also important to remark that other properties such as stellar rotation can heavily influence the evolutionary paths followed by massive stars.
Moreover, the presence of a companion star can also have a significant impact on the evolution of massive stars. In fact, binary interaction can lead to mass transfer, mass accretion, common envelope evolution, and even stellar mergers \citep[see][and references therein]{marchant23}. Some of these processes can trigger a ``rejuvenation'' of the binary product, where the star may become hotter and more luminous as a result of the additional hydrogen available for fusion. 

The initial mass of the star is the main parameter that determines the duration of the main sequence phase; however, other factors can also play an important role in this phase \citep{langer12}. One of them involves the internal mixing mechanisms that can provide the core with fresh hydrogen, such as rotation, core overshooting, inflation, or semi-convection \citep[see][]{maeder88, maeder09, schootemeijer19, martinet21}. However, these mechanisms are far from being well understood and are hard to test empirically.
Among them, the effect of rotation has been extensively studied as it can be confronted with observations. In this regard, the impact on evolution highly depends on the initial rotation rate of the star when it enters the ZAMS, and whether this rotation is maintained or not during the main sequence \citep[e.g.,][]{meynet05, heger05, ekstrom12, georgy13}.
Moreover, rotation can also enhance mass loss through stellar winds, which is another factor that can have a significant impact on the evolution of massive stars (see below). Unfortunately, as indicated before, the initial rotation rates of massive stars are not yet well constrained \citep[see, however,][]{holgado22}, preventing us from adopting realistic initial conditions. 

Mass loss also has an important impact on the main sequence of massive stars \citep[e.g.,][]{kudritzki00, vink10, smith14, Renzo17}. As indicated in Sect.~\ref{intro.1.1.importance}, these objects are characterized by strong stellar winds, which are able to ``unpeel" the outer layers of the star and take away angular momentum from the surface. Furthermore, the mass loss rate of massive stars is not necessarily constant, but it can vary significantly during the main sequence phase. In this regard, changes in the opacity of the atmosphere along the evolution can also lead to enhanced mass-loss rates. These rates can also be severely affected by the presence of magnetic fields, which can lead to the formation of magnetically confined winds \citep[e.g.,][]{Petit17, keszthelyi19}.
In fact, the effects of magnetic fields can also have a significant impact on the evolution of massive stars on the main sequence \citep{keszthelyi17, keszthelyi19, keszthelyi22}.

Last but not least, the metallicity of the star, which is defined as the fraction of elements heavier than helium in its atmosphere, also plays an important role in the evolution of massive stars on the main sequence \citep[see][]{kudritzki87, vink01}. In fact, the higher the metallicity of the star, the stronger the metal lines that drive the stellar winds, directly affecting the amount of mass loss \citep{mokiem07}. Furthermore, metallicity influences the initial amount of CNO material in the nuclei, making stars more compact towards lower metallicities \citep{yoon06, ekstrom12, szecsi15}.

\subsection{The post-main sequence phase and final fates}
\label{intro.1.4.post-ms}

The post-main sequence evolution has much shorter timescales compared to the main sequence, and it is characterized by a series of rapid changes in the structure and properties of the star. At the beginning of this phase, the star has an inert helium core surrounded by a hydrogen burning shell. The lack of energy production leads gravity to compress the core until its temperature and pressure are high enough to ignite the helium burning. The large amount of energy produced in the hydrogen shell and the core leads to an expansion of the outer layers of the star, which become significantly cooler and more luminous. The stars reaching this phase become RSGs\footnote{Transitioning objects may include the so-called ``yellow supergiants" (YSGs).}, with effective temperatures between 4000\,K and 3000\,K \citep[K-to-M spectral types;][]{levesque05} and radii of some hundreds of solar radii (see Fig.~\ref{fig:HR_Marchant}). As the star evolves, the helium in the core is also exhausted. In contrast to the hydrogen burning phase, the helium burning phase is much faster and only last for 0.3\,--\,2 million years for stars born with 12\,--\,120\,M$_{\odot}$ \citep[see][]{ekstrom12}. At this point, a main difference with the less massive stars is that the core of the star is able to reach temperatures high enough to ignite heavier elements, such as carbon, neon, oxygen and silicon, each time producing a burning shell around the core beneath the previous burning shell \citep[see][]{laplace21}. It is during the RSG phase that the star undergoes several dredge-up episodes, where the outer layers of the star are mixed with the material from the core, changing the surface abundances of the star \citep{heger10}. Moreover, this phase is characterized by an important mass loss of the outer layers of the stars \citep{humphreys20}. It is also predicted in some cases that stars experience an evolution towards hotter temperatures through the so-called ``blue-loops'' \citep{ekstrom12}. Regardless of these loops, the fusion of heavier elements will also lead to the formation of an iron core, which, rather than providing energy to the star through its fusion, absorbs it in an endothermic process. At this point, this iron core is surrounded by a series layers of other elements in an ``onion-like'' structure. The core contraction will continue until the temperature eventually reaches $\sim$10$^{10}$\,K. Stars born with M \gs8\,M$_{\odot}$ have reached the last phase of their evolution \citep{woosley86}.

In a process that lasts for a few seconds, the photons are so energetic that they are able to photo-disintegrate the iron core into $\alpha$ particles, reducing the radiative pressure, leading to the catastrophic collapse of the star and a subsequent supernova explosion \citep{janka96, janka12}. During this event, vast amounts of high-energetic particles (such as $\gamma$ rays or neutrinos) are ejected into space. Furthermore, different heavier elements are produced, which are expelled into the interstellar medium with the outer layers of the star, enriching the interstellar medium (see again Fig.~\ref{fig:elem}). At the moment of collapse, the core of the star reaches nuclear matter densities of 10$^{14}$\,g\,cm$^{-3}$. A degenerate neutron star is formed, which is held by the neutron degeneracy pressure \citep{woosley02, steiner10}. It is believed that massive stars with initial masses above a given threshold surpass the Tolman-Oppenheimer-Volkoff limit, and the neutron star collapses into a black hole \citep{oconnor11, Sukhbold16}.

\section{Context of the thesis}
\label{intro.2.context}

From the different evolutionary phases of massive stars, the one in which they transition from the main sequence towards the RSG phase is among the most elusive, less constrained, and yet important. Understanding the properties of stars in this transition phase is crucial for testing some of the predictions of stellar evolution models and for setting strong anchor points to our overall theoretical knowledge of massive stars and their end products. Back in the decade of the 1980s, it was generally accepted that main-sequence massive stars comprise either O-type stars or early B-type giants to dwarfs. B-type supergiants and bright giants were thought to be transitioning objects between the main sequence and the RSG phase, this is, post-MS objects. Furthermore, the number of dedicated studies to these objects before the 1990s was limited to a limited number of works \citep[e.g.,][]{cayrel58, jaschek67, walborn70, dufton72, melnick85, lennon86}. However, with the advent of CCD photometry and slit spectroscopy, the number of observed massive stars with available stellar parameters increased significantly. The works by \citet[][]{fitzpatrick90} in the Large Magellanic Cloud (LMC), the ones by \citet{lennon92, lennon93} in the Milky Way, or the one by \citet{blaha89} including stars from both groups, were some of the pioneering works that stepped up the number of studied B-type supergiants. The former led to the realization of the unexpectedly large population of these stars in the Hertzsprung--Russell (HR) diagram \citep[see also][]{humphreys84, garmany89}.

\begin{figure}
    \centering
    \includegraphics[width=0.8\linewidth]{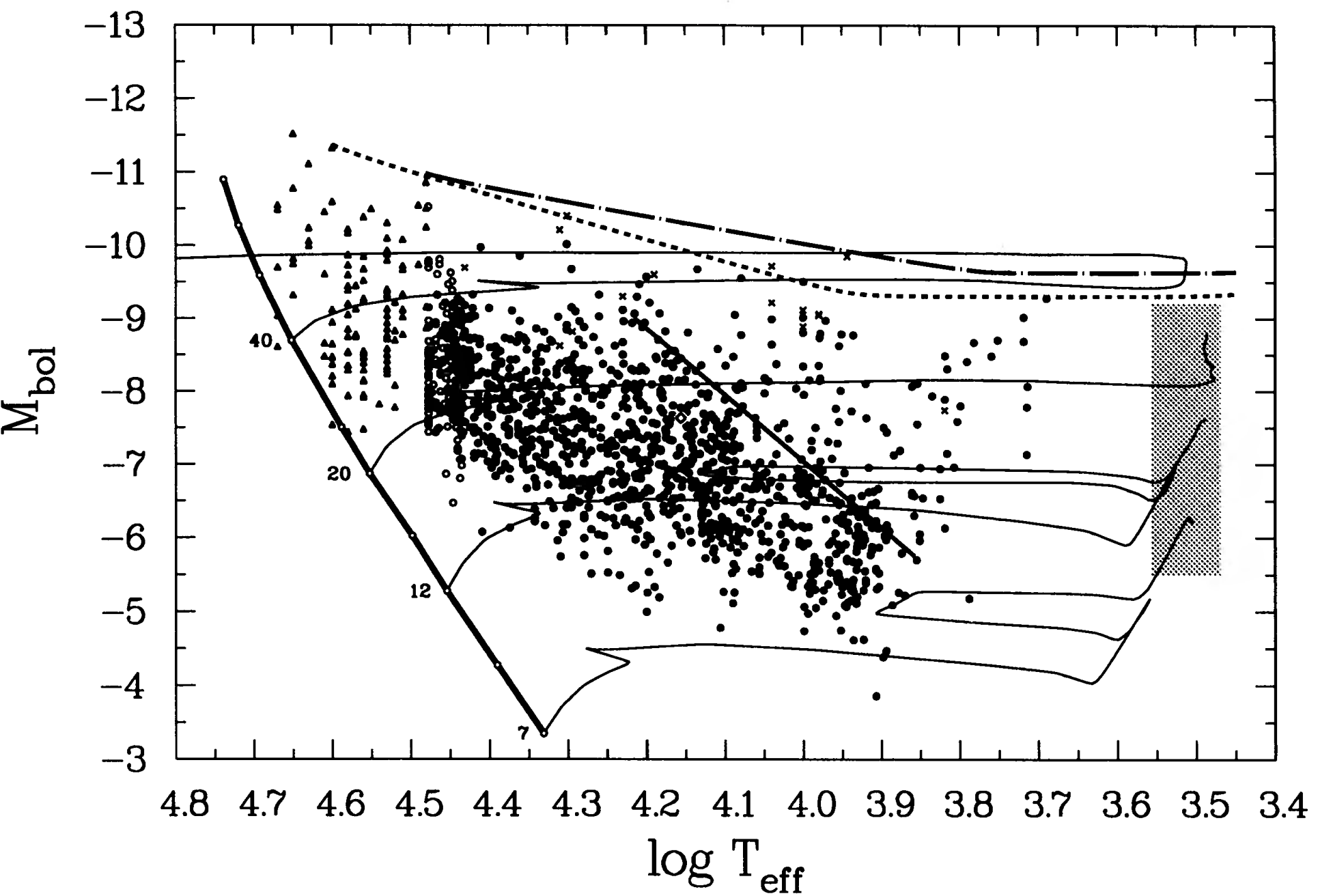}
    \caption{HR diagram for the LMC taken from \citet[][\textcopyright AAS, reproduced with permission]{fitzpatrick90}.
    The location of the ``ledge" is shown by the solid diagonal line. The heavy solid line shows the location of the ZAMS for models published by \citet{maeder88}. Thin solid lines are evolutionary tracks for four different masses.}
    \label{fig:HR-fitz}
\end{figure}

The study by \citet{fitzpatrick90} was based on a sample of 1375 O-to-G-type supergiants that only included objects that do not exhibit optical emission lines (i.e., did not include WR or LBV stars). It was found a diagonal ``ledge" of B-type supergiants where the density of objects dropped over five times from the hot to the cool side of the diagram (see Fig.~\ref{fig:HR-fitz}). Furthermore, the study revealed that the density contrast across the ledge was unlikely to result from random chance. The comparison with evolutionary tracks by \citet{maeder88, maeder89} revealed that the ledge was located 5 to 10\,kK beyond the terminal-age main sequence (TAMS) of the models\footnote{\citet{langer92} in fact defined this problem as the ``TAMS problem".}. 
This led to the interpretation that all of those B-type supergiants were in a post-MS phase, thus no longer in the core hydrogen-burning phase. Moreover, further checks for different masses and metallicities indicated that the drop should be almost vertical when the effective temperature decreases, in contrast to the diagonal shape of the ledge. The most plausible explanation for \citet{fitzpatrick90} was found if massive stars follow a blue-loop evolution \citep[see also][]{chiosi70,  paczynski70}. In this regard, \citet{simpson71} was the first to suggest that, compared to main sequence stars, stars that have undergone a RSG evolution should exhibit enhanced helium and nitrogen surface abundances as a result of a deep surface convection zone operating during the RSG phase. However, predicted mass-loss rates back then were not sufficient to ``unpeel" the stars in order to expose the inner layers of processed material. The data of \citet{fitzpatrick90} also evidenced the lack of a gap in the HR diagram caused by the rapid evolution of stars leaving the main sequence. In this regard, \citet{tuchman90} suggested that the gap was filled by stars that have accreted helium from a companion. This put our understanding of the evolutionary nature of early to mid-B-type supergiants under debate, opening a discussion that has remained open until the present. 

Despite the fact that the discrepancy with the evolutionary models was worrying, it was not the only existing one involving B-type supergiants.
The ``mass discrepancy problem" was another important issue that was not (and still is not) resolved. First pointed out by \citet{groenewegen89} and \citet{herrero90}, it arose from the discrepancy between the stellar masses derived from the spectroscopic analyses with adopted distances and the masses obtained from the evolutionary tracks, the second ones being systematically larger than the first ones by up to 50\% \citep[see also][]{kudritzki92, herrero92}.
An additional problem was also found from the non-negligible fraction of luminous blue stars on the main sequence that exhibited enhanced surface helium abundances \citep[e.g.,][]{schonberner88, voels89, gies92, herrero92} where evolutionary models did not predict such enhancements. In this regard, \citet{langer92} proposed rotationally induced mixing as a possible solution to this problem. However, this implied that helium-enriched stars should be fast-rotating objects, that helium enrichment must be accompanied by nitrogen enrichment, or that all B-type supergiants within the theoretical post-main sequence gap should be helium enriched.

Parallel to the earliest works on B-type supergiants and the above-mentioned findings was the theoretical development of the radiative transfer theory required for the analyses of hot massive stars \citep[e.g.,][]{sobolev57, hummer68, mihalas65, castor79, pauldrach86}, including the \citet*[][CAK]{castor75} theory of radiation-driven winds. The works of \citet{dufton72, dufton79}, which used local thermal equilibrium (LTE) model atmospheres to analyze two early B-type supergiants, were followed by those of \citet[][]{lennon93} and \citet[][]{mcerlean99}, which employed plane-parallel hydrostatic non-LTE (NLTE) models for the analyses. Additional advancements followed with the inclusion of line-blanketing effects in the model atmospheres \citep[e.g.,][]{vacca96, hillier98}, the presence of wind inhomogeneities \citep[e.g.,][]{schmutz95, hamann98}, or the theoretical improvement of the H$\alpha$ formation \citep{puls96}.

Naturally, these steps also led to the development or improvement of the model atoms required to compute the line formation and synthetic spectra. These first included hydrogen and helium, but followed the most abundant metals in the atmospheres of massive stars such as carbon, nitrogen, oxygen, silicon, and iron group elements, which are important for the blanketing effect \citep[][]{jokuthy02, eber88, becker88, becker89, herrero92}.\\

The last two and a half decades represent a turning point in the study of massive stars at most levels. The theoretical developments on the physical processes that take place in the stellar interior, atmospheres, and winds of massive stars \citep{kudritzki00, maeder00}, led to a high level of confidence and reliability in the output of the stellar atmosphere and stellar evolution codes \citep[see,][respectively]{puls03, martins13}. Two key aspects in the evolution of massive stars such as stellar rotation and mass loss were taken into account in the models, explaining many of the observed properties of massive stars in the first decade of the 21st century. However, other aspects of the evolution of massive stars have remained unsolved.

The new millennium brought new challenges to the field of massive stars and B-type supergiants. The advent of new observational facilities provided a wealth of new data of much higher resolving power and spectral coverage with respect to those used before, which allowed for a more detailed analysis of the spectral features and surface abundances of the stars \citep[e.g.,][]{webmayer22, webmayer23}. Some of the most relevant works in this regard were those by \citet{crowther06, lefever07, markova08, searle08}. Their results revealed that the predictions of the state-of-the-art models \citep[such as the widely used models by][]{brott11a, ekstrom12} were still unable to reproduce some of the observed properties of massive stars. For instance, the observed surface abundances revealed a mismatch between observations and the predictions by single-star evolutionary models regarding the efficiency of rotational mixing during the main sequence \citep{hunter08, brott11b, rivero-gonzalez12}. Furthermore, different mechanisms of angular momentum transport from the core to the surface \citep[e.g.,][]{brott11a, ekstrom12} or the spin-down effect of magnetic braking \citep{meynet11, keszthelyi20} were used to partially describe the observed rotation rates of massive stars, which were found in the form of a bimodal distribution \citep{conti77, ramirez-agudelo13, dufton13, ramirez-agudelo15, holgado22}. Another apparent discrepancy was found between the theoretically predicted and observationally derived mass-loss rates, which have a significant impact on the outcome of stellar evolution computations \citep{keszthelyi17}. In parallel, the long-standing suspicion about the high percentage of massive stars born in binary and multiple systems was finally empirically confirmed in the O- and main-sequence B-type star domains \citep[see][for recent reviews; see also Fig.~\ref{fig:HR_Marchant}]{vanbeveren17, barba17, sana17}. B-type supergiants, which were once thought to be single stars \citep{humphreys78}, were now expected to include a significant fraction of binary products \citep{demink14, farrell19}. 

The last decade has been marked by the development of new theoretical works that describe the evolution and outcomes of the different types of binary interaction in massive stars \citep[e.g.,][]{langer20}. On the observational side, many efforts have been made to try to identify binary systems and signatures of past or ongoing binary interaction. These include the search for spectroscopic binaries, surface abundance signatures of merging or mass-transfer processes, or the presence of circumstellar material. In this regard, the O-type stars domain has been the most studied in the search for binary systems \citep[e.g.,][]{manson09, barba10, sana13, maiz-apellaniz19}, whereas the B-type supergiants domain is less constrained \citep{mcevoy15, menon24}. Given this context, observational clues about the evolutionary status of B-type supergiants represent a key piece of the puzzle to understand the properties of massive stars, which brings us back to the original work of \citet{fitzpatrick90} on the overdensity of these objects beyond the main sequence.

A recent development in the study of massive stars comes from the advent of asteroseismology \citep[see][]{aerts21}. This technique has been successfully applied to intermediate- and low-mass stars, providing a wealth of information about their internal structure and evolution \citep[see, e.g.,][]{Chaplin13, garciaRA19}. However, the application of asteroseismology to massive stars was initially constrained by the lack of the necessary data sets \citep[see some early works on B-type stars by][]{aerts03, dupret04}. This situation improved with the launch of the \textit{CoRoT} space mission \citep[e.g.,][]{Degroote10, Blomme11, Neiner12}, and, particularly in the last decade, with the advent of new observational facilities such as the \textit{Kepler/K2} \citep{koch10, borucki10, howell14} and TESS \citep{ricker15} space missions. In particular, the latter has obtained photometric light curves for the largest number of massive stars to date, with high photometric precision, temporal cadence, and coverage. These data have revealed that massive stars are dominated by stochastic low-frequency variability (SLFV), which origin is yet to be explained (see \citeauthor{Buysschaert15} \citeyear{Buysschaert15}, \citeauthor{David-Uraz17} \citeyear{David-Uraz17}, \citeauthor{bowman20b} \citeyear{bowman20b}; also \citeauthor{bowman19a} \citeyear{bowman19a} as the ``discovery” paper).

However, one of the main challenges comes from the difficulty of separating the oscillations (or pulsation-mode frequencies) produced in the stellar surface or stellar interiors from the variability induced by the stellar winds. 
In this regard, combining information from photometric and spectroscopic variability provides further insights into the internal structure of massive stars \citep[see the works by][]{aerts18, burssens23}.

Despite the challenges, asteroseismology represents a powerful tool to investigate some of the physical processes that take place in the core of massive stars, such as the mixing of nuclear products, the transport of angular momentum, and the effects of magnetic fields. This information is crucial for testing the predictions of stellar evolution models and for understanding the properties of massive stars in general \citep{aerts19, bowman20a}.

Another important milestone from recent years comes from the ESA \textit{Gaia} mission \citep{gaiacollaboration16, gaiacollaboration23, babusiaux23}. \textit{Gaia} has provided valuable and reliable information about the distances and proper motions of a very large fraction of massive stars in the Milky Way. This information is crucial for accurately determining the physical fundamental properties of massive stars (such as their luminosities, radii, and masses) but also to study their kinematics relative to their parent stellar clusters or associations. The latter is crucial to find runaway objects that have been ejected as a result of, for example, binary interaction or supernova kick \citep{gvaramadze11, perets12, renzo19}.

At present, we live in a bright time for the study of massive stars. The availability of large databases of high-quality spectroscopic, photometric, and kinematic data combined with the recent theoretical developments for the physical processes taking place in the stellar interiors, atmospheres, and winds of massive stars \citep[e.g.,][]{aerts19, marchant23, vink23}, as well as in the binary interaction processes \citep[e.g.,][]{schneider19, renzo21, sen22}, has provided a unique opportunity to study massive stars at unprecedented detail. This situation also provides a unique opportunity to address some of the long-standing questions involving B-type supergiants, test the predictions of the state-of-the-art evolutionary models, and provide new insights into their evolutionary nature.

\section{The IACOB project}
\label{intro.3.iacob}

In order to improve the situation described above, in 2008 was formally born the long-term observational project IACOB \citep[see][]{simon-diaz11a, simon-diaz15, simon-diaz20}. The main objective of this ambitious project hosted at the Instituto de Astrofísica de Canarias, is to provide a statistically significant empirical overview of the main physical properties (including spectroscopic parameters and abundances) of O and B-type stars in the Milky Way. The ultimate goal is to provide reliable anchor points for theoretical models of stellar atmospheres, winds, interiors, and evolution of massive stars. To achieve this, the IACOB project has compiled a unique database of high-resolution, multi-epoch optical spectra of Galactic O- B- and A-type stars over the last 15 years. Currently, this database comprises an outstanding number of more than 15\,000 spectra from more than 2000 sources. The two main observing instruments are the FIES \citep{telting14} and HERMES \citep{raskin11} fiber-fed echelle spectrographs attached to the 2.56\,m Nordic Optical Telescope (NOT) and the 1.2\,m Mercator telescope, respectively, both located at the Roque de los Muchachos Observatory in La Palma, Spain (see Chapter~\ref{chapter1} for more details).

Three key properties of the IACOB project are: (1) the quality of the data, which have been obtained using advanced and stable high-resolution echelle spectrographs, allowing for a more detailed analysis of the spectral features of massive stars \citep[e.g.,][]{simon-diaz14a, godart17, simon-diaz17}; (2) the analysis tools developed by the IACOB team and collaborators, which allow for a detailed quantitative spectroscopic analysis of the observed stars \citep[e.g.,][and Chapter~\ref{chapter2}]{simon-diaz14b, holgado18}; and (3) the collection of multi-epoch data for a significant fraction of the observed stars, which allows for the study of the variability of massive stars \citep{burssens20, britavskiy23}.

The IACOB project has also established important synergies with other large-scale studies of massive stars. Some of them in the Large and Small Magellanic Clouds include the VLT-FLAMES Tarantula Survey \citep[VFTS;][]{evans11, evans15}, the VLT-FLAMES Survey of Massive Stars \citep[VFMS - P.I. S. Smartt; e.g.,][]{evans08, hunter07, trundle07, lennon22}. Others in the Milky Way include the Galactic O-Star Spectroscopic Survey \citep[GOSSS;][]{sota11, sota14, maiz-apellaniz16, arias16}, or the Southern Galactic O- and WN-type stars \citep[OWN;][]{barba10, barba14, barba17}. Moreover, the distances and visual magnitudes covered by the IACOB sample perfectly overlap with the brightest sources included in the upcoming spectroscopic surveys such as WEAVE-SCIP-OB \citep{dalton20, jin24} and 4MOST-4MIDABLE-LR \citep{chiappini19} in the Northern and Southern hemispheres, respectively.\\

\section{Aim and structure of the thesis}
\label{intro.4.aimstructure}

\subsection{Aim of this thesis}
\label{intro.4.1.aimPhD}

This thesis work sets its bases in the framework of the IACOB project. It is conceived as an empirical reappraisal of the properties of B-type supergiants in the Milky Way, aiming at providing new insights into the evolutionary status of these objects, which are key to understanding the overall properties of massive stars.
To achieve this, this work has benefited from a unique database of a growing number of high-resolution, multi-epoch optical spectra obtained by the IACOB team. 
Second, this work used semiautomatic tools to perform the detailed quantitative spectroscopic analysis of the observed stars, which represented a significant improvement compared to traditional techniques (see Chapter~\ref{chapter2}). In particular, this has allowed to obtain estimates of the broadening, stellar, and wind parameters, and surface abundances of key atomic elements.
Third, multi-epoch data have been used to study the variability and multiplicity of the observed stars.
Finally, the work has been complemented with additional information on parallaxes, proper motions, and photometric variability provided by the \textit{Gaia} and TESS missions.

All these aspects together make this work a significant improvement with respect to the previous studies on B-type supergiants, which were based on low-resolution spectra, a reduced number of sources, or were biased in the sample selection or limited to objects in the Magellanic clouds or nearby galaxies. The reference work of this thesis in terms of sample size is the one by \citet{castro14}, which compiled results for 439 Galactic OB stars observed spectroscopically (see Chapter~\ref{chapter4}).
Furthermore, this thesis represents an important achievement within the IACOB project. While the Ph.D. thesis of \citet{holgado19} focused on the study of the properties of 415 O-type stars of all luminosity classes, this one has concentrated in their direct descendants, these are, the B-type supergiants. \\

The ultimate goal of this thesis is to provide new empirical constraints to the state-of-the-art evolutionary models, and add new clues towards nature of the B-type supergiants included in the sample. In addition to that, this study aims to have some impact across several fields of astrophysics. In fact, supernovae, WR stars, RSGs, stellar black holes, neutron stars, long-duration gamma-ray bursts, are all direct descendants of massive OB stars. Any meaningful attempt to connect these extreme objects with their progenitors is critically supported by our knowledge of massive star evolution. Stellar population synthesis codes and the interpretation of the light emitted by starburst galaxies also rely on a reliable characterization of the physical and evolutionary properties of massive stars.

\subsection{Building the spectroscopic sample}
\label{intro.4.2.sample}

At the time of the start of this thesis, the IACOB project had access to 8000 spectra of about 1000 stars, half of which corresponded to O-type stars. The other half was distributed among B- A- and M-type stars or unclassified sources. In contrast with the O-type stars, for which different systematic projects had provided a high level of completeness and a homogeneous spectral reclassification (the GOSSS survey) or have studied their multiplicity (the OWN and MONOS\footnote{See \citet{maiz-apellaniz19, trigueros-paez21}.} surveys), the B-type stars compiled by the IACOB project lacked from any of these aspects. The sample of B-type stars was built from the compilation of different observing campaigns carried out by the IACOB team and collaborators throughout the years aimed at studying specific groups of stars or regions of the Galaxy. Therefore, a significant work was initially required to review and homogenize the sample. In this regard, a first step consisted in the manual revision of about 4000 spectra to check for potential issues in the data and metadata (e.g., issues with the telescope pointing, wrong header identifiers, problems in the wavelength calibration, among many others). This was followed by the development of a semi-automatic tool to perform the identification of the potential B-type supergiants in the sample regardless of their quoted classification, which resulted in a total of 400 stars. 

In parallel, several observing proposals to the Spanish Time Allocation Committee (CAT) were prepared, which sought to improve the homogeneity of the sample of B-type supergiants and reduce observational biases of the IACOB spectroscopic sample in the northern hemisphere. These resulted in two successful observing campaigns at the NOT telescopes (36-NOT5/20A and 59-NOT7/20B, P.I. A.~de Burgos), and a very successful Large-Program for the NOT and Mercator telescopes (P.I. S.~Simón-Díaz, co-P.I. A.~de Burgos). The latter was awarded with three years of several observing campaigns. All of these campaigns significantly improved the number of observed B-type supergiants during this thesis.

The observations were complemented with the exploitation of the European Southern Observatory (ESO) data archive. In particular, retrieving all available spectra of OB stars taken with the FEROS high-resolution Echelle instrument \citep{kaufer97}, attached to the MPG/ESO 2.2\,m telescope at La Silla observatory, Chile. Thanks to the several successful preparation and completion of the observing runs, and to the publicly available FEROS data, the final sample of B-type supergiants used for this thesis included 900 sources by the end of 2023.

\subsection{Structure of this thesis}
\label{intro.4.3.structure}

The structure of this thesis follows the logical steps of any empirical study, namely: (1) compilation of the data and presentation of the working sample; (2) analysis of the data and presentation of the main results; and (3) implications of the results relevant to the field of study.

Given the large volume of data contained in this thesis, the previous steps were separated into different blocks, each of them aiming to result in a publication in a peer-reviewed journal. Therefore, this thesis is structured as a compendium of articles, each of them focused on a specific aspect of the study of B-type supergiants. At the time of thesis submission, three articles have been accepted or published, comprising Chapters~\ref{chapter1}, \ref{chapter2} and \ref{chapter3} of this thesis. Two more articles are in the last stage of preparation and are expected to be submitted for publication within the next few months. They comprise Chapters~\ref{chapter4} and \ref{chapter5}.

The first article (Chapter~\ref{chapter1}) presents the compilation and selection criteria of the spectroscopic sample used in this thesis, which initially comprised ~750 blue luminous stars with spectral types O9 to B9, but was later expanded thanks to new observation of missing stars. It also included a first completeness analysis, the identification of double-line spectroscopic binaries (SB2), which were later used in the last article, and the adoption of distances to all the stars, which were used in the last three articles.

The second article (Chapter~\ref{chapter2}) presents the methodology used for the quantitative spectroscopic analysis of the sample, including the determination of the line-broadening, spectroscopic and wind parameters, and the surface abundances of key elements: helium and silicon. These results opened up different lines of study related to each of the parameters derived, which are connected with different open questions in the evolution of massive stars.

In this last regard, one of the key questions related to the wind properties involves the actual mass-loss rates of massive stars, which have a significant impact in their evolution \citep[see][]{keszthelyi17}. In fact, the widely used prescriptions of the rates by \citet{vink01} predict an important increase in the mass-loss rate when B-type supergiants cross the so-called bi-stability jump towards cooler temperatures. This has been used to explain the lack of fast-rotating objects beyond the jump \citep{vink10}, but this lack may also outline the termination of the main sequence \citep{brott11a}. Motivated by previous observational evidence that the jump may not be present, and by the visit of Dr. Z. Keszthelyi to the IAC in November\,-\,December 2023, the third article (Chapter~\ref{chapter3}) set the focus of the study on evaluating whether the jump in mass-loss rates predicted by \citet{vink01} was present or not.

The lack of detection of the jump in the mass-loss rates led to investigate the location of the termination of the main sequence of massive stars. The fourth article (Chapter~\ref{chapter4}) can be interpreted as the continuation of the works led by \citet{fitzpatrick90} and \citet{castro14} for which, in addition to using the largest and most homogeneous spectroscopic sample of B-type supergiants compiled to date, accounts for two additional properties to evaluate the main sequence end. These are the rotational properties and the distribution of single-line spectroscopic binaries (SB1) across the HR diagram. Ultimately, the results of this article allow to provide new constraints to the evolutionary models in the 12\,--\,40\,M$_{\odot}$ mass range.

The fifth and last article (Chapter~\ref{chapter5}) brings all the previous results together, offering new insights into the nature of the B-type supergiants in the sample. 
Alongside their rotational properties, helium surface abundances, and multiplicity, it compares derived spectroscopic masses with evolutionary ones to find potential post-RSGs and binary interaction products. Furthermore, it includes preliminary analysis of CNO surface abundances and explores the pulsational properties of a sub-sample of stars for which photometric data from TESS is available.\\

The last chapter of this thesis (Chapter~\ref{chapter6}) summarizes the conclusions of the different articles. It also provides future research lines that will surely follow the results presented in this thesis. The latter especially considering the arrival of the WEAVE-SCIP-OB and 4MOST-4MIDABLE-LR spectroscopic surveys, which will increase the number of observed B-type supergiants in the Milky Way by several thousands.

\section{Addendum}
\label{intro.5.addendum}

The research work of this thesis has been complemented by different activities and fruitful collaborations that were beyond its original scope and positively enriched its development and content. Some of them include:

\begin{itemize}

    \item Plan and carry out the observations for a total of 78 nights at the Roque de los Muchachos observatory using the NOT and the Mercator telescopes.

    \item Perform two stays at the University of Innsbruck, Austria, to collaborate with Dr.~Miguel A. Urbaneja (co-supervisor of the thesis) in the analysis of the sample stars.

    \item Publish an additional study on the massive stellar population in the Perseus OB1 association \citep{deburgos20}. This research began under the IAC Summer Grants program for astronomical research and is the foundation of the research project for the recently awarded ESO Fellowship in Chile, starting in November 2024.

    \item Publish additional works in peer-reviewed journals, some of which are closely connected with the research line of this thesis \citep{burssens20, lennon21, negueruela23, negueruela24}, others which resulted the participation in the WEAVE-SCIP survey \citep[][]{Monguio20, shoko24}, and some others from other fields of astrophysics \citep[][]{Heintz20, vaduvescu22, domcek23}.

    \item Develop two open-source Python packages available to the community:
``\href{https://github.com/Abelink23/pyIACOB}{\textit{pyIACOB}}" (see Appendix~\ref{apdx.pyIACOB}), which aims at users of the IACOB spectroscopic database for analyzing and performing different measurements on the spectra; and ``\href{https://github.com/Abelink23/LCExtractor}{\textit{LCExtractor}}", which is used to extract and reduce the TESS and \textit{Kepler/K2} light curves, and to obtain the associated periodograms.

\end{itemize}

\clearemptydoublepage{}

%
%
\chapter{Building a modern empirical database of Galactic O9\,--\,B9 supergiants}
\label{chapter1}
{\flushright{\it All we have to decide is what to do \\with the time that is given us.\\\smallskip}{\rm \small J.R.R. Tolkien, The Fellowship of the Ring\\}}

\vspace{0.7cm}
%
%

\begin{center}
This chapter includes the content of the paper accepted in\\ Astronomy and Astrophysics under the reference: \href{https://doi.org/10.1051/0004-6361/202346179}{aa46179-23}.    
\end{center}

\includepdf[pages=-]{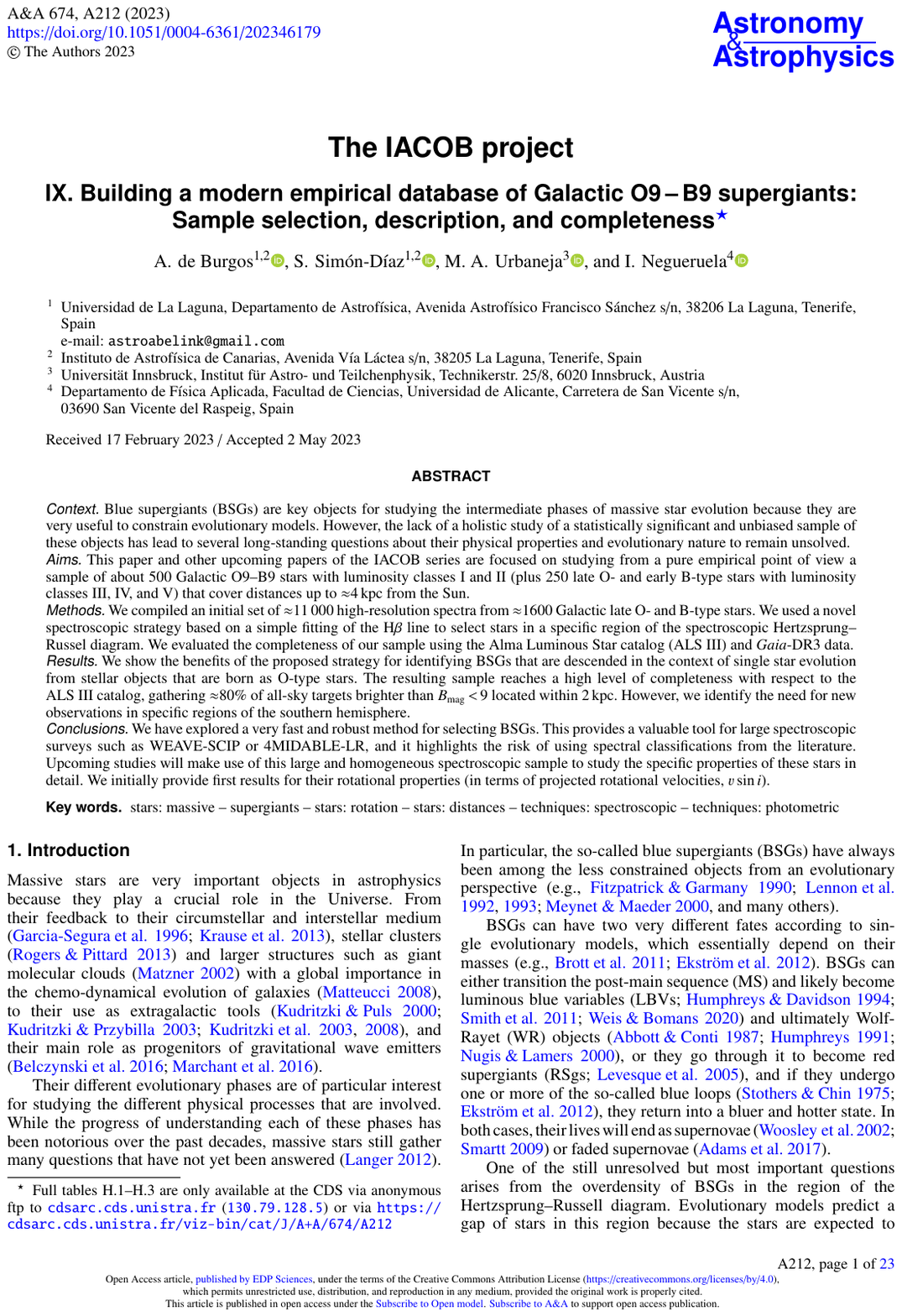}

\clearemptydoublepage{}

%
%
\chapter{Large-scale quantitative spectroscopic analysis of Galactic luminous blue stars}
\label{chapter2}
{\flushright{\it I would prefer to stay up and \\watch the stars than go to sleep\\\smallskip}
            {\rm \small Vera Rubin \\}}

\vspace{0.7cm}
%
%
\begin{center}
This chapter includes the content of the paper accepted in\\ Astronomy and Astrophysics under reference: \href{https://doi.org/10.1051/0004-6361/202348808}{aa48808-23}.    
\end{center}
\includepdf[pages=-]{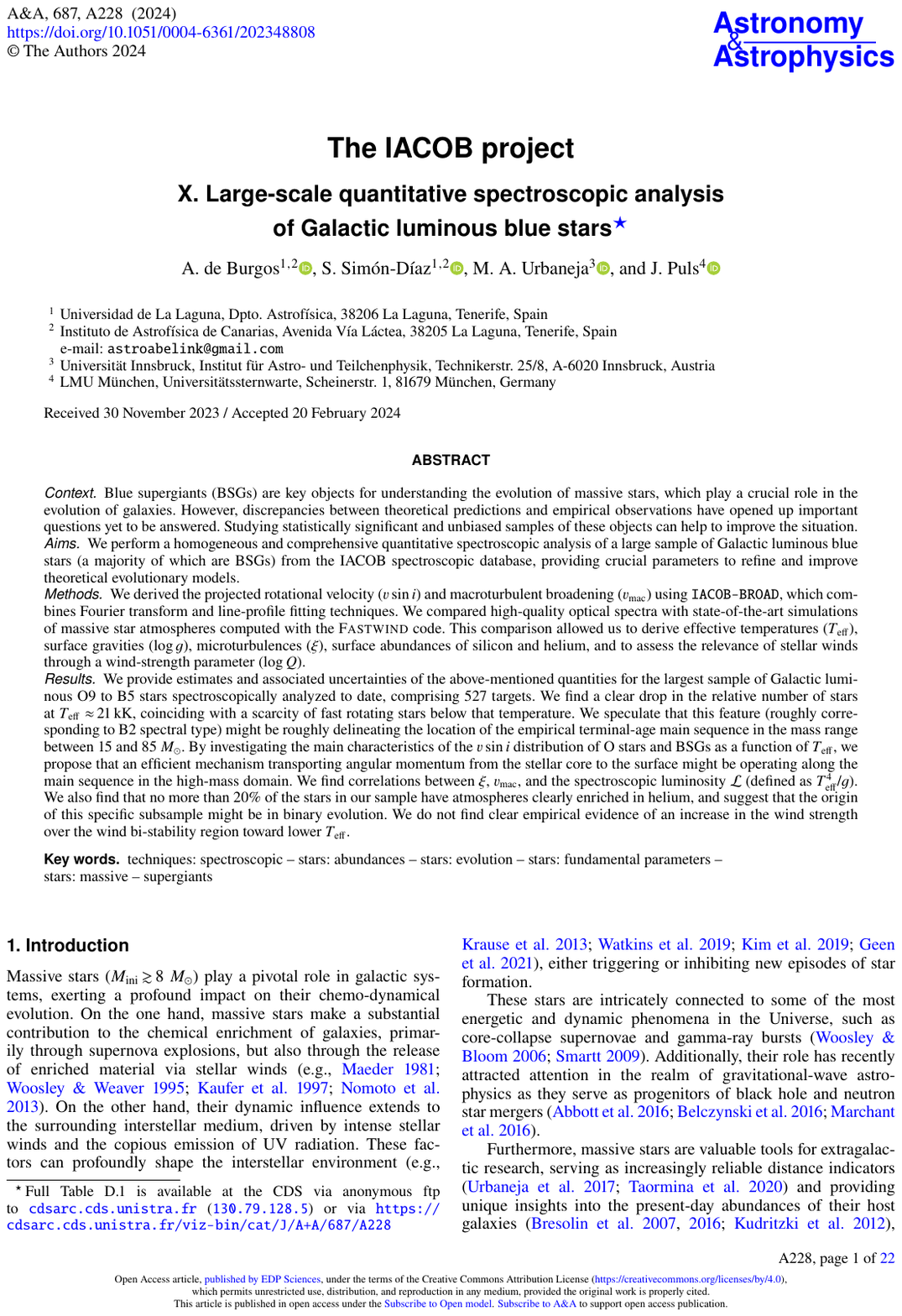}

\clearemptydoublepage{}

%
%
\chapter{No increase of mass-loss rates over the bi-stability region}
\label{chapter3}
{\flushright{\it The most subversive people are\\
                  those who ask questions.\\\smallskip}
            {\rm \small Jostein Gaarder, Sophie’s World\\}}

\vspace{0.7cm}
%
%
\begin{center}
This chapter includes the content of the paper accepted in\\ Astronomy and Astrophysics under the reference: \href{https://doi.org/10.1051/0004-6361/202450301}{aa50301-24}.    
\end{center}
\includepdf[pages=-]{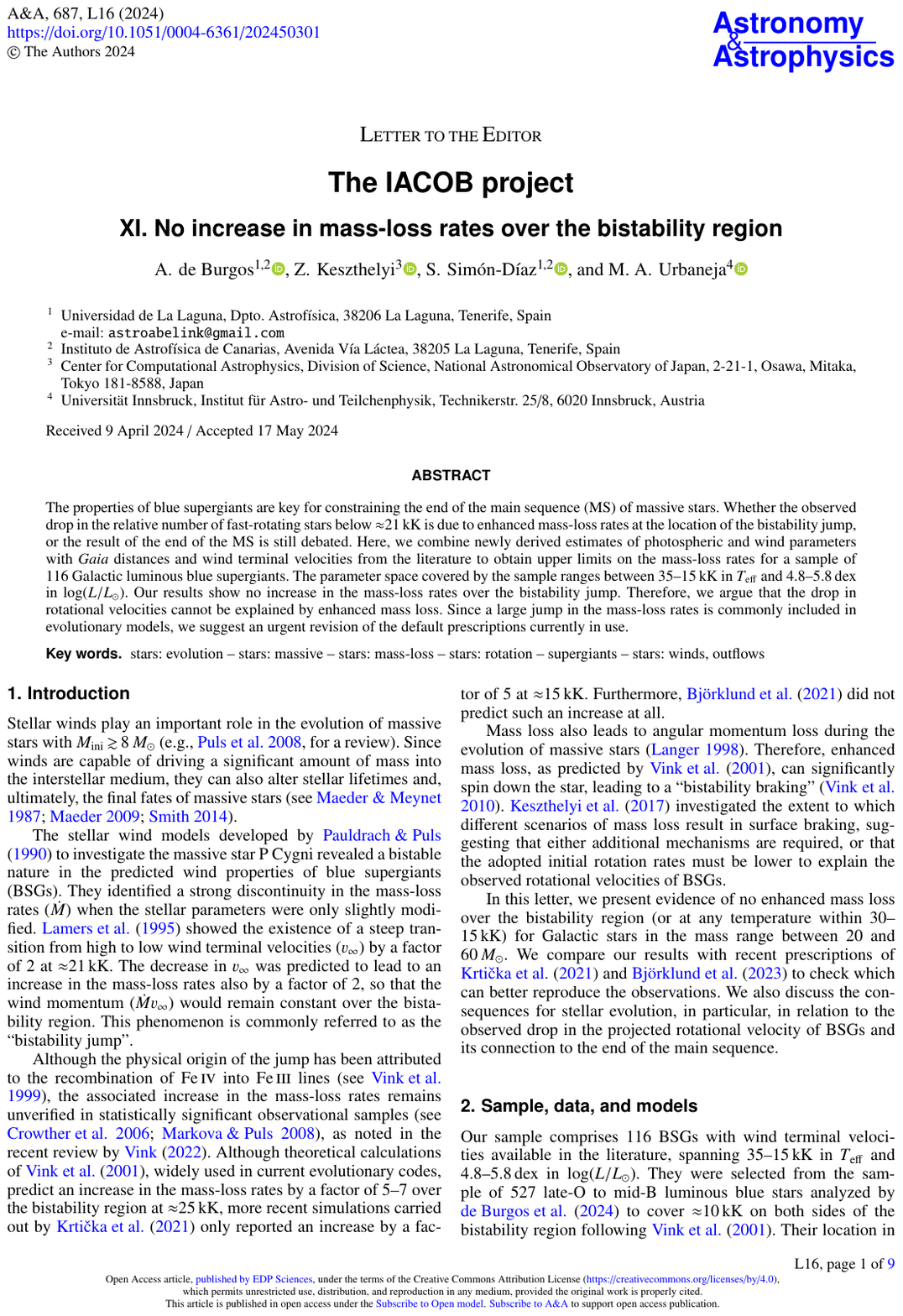}

\clearemptydoublepage{}

%
%
\chapter{New clues on the location of the TAMS in the massive star domain}
\label{chapter4}
{\flushright{\it Hoping for the best, prepared for the worst, \\
                  and unsurprised by anything in between.\\\smallskip
            }{\rm \small Maya Angelou\\}}

\vspace{0.7cm}
%
%
\begin{center}
This chapter includes the content of a paper in the last stage\\ of preparation that will be submitted to Astronomy and Astrophysics.
\end{center}
\includepdf[pages=-]{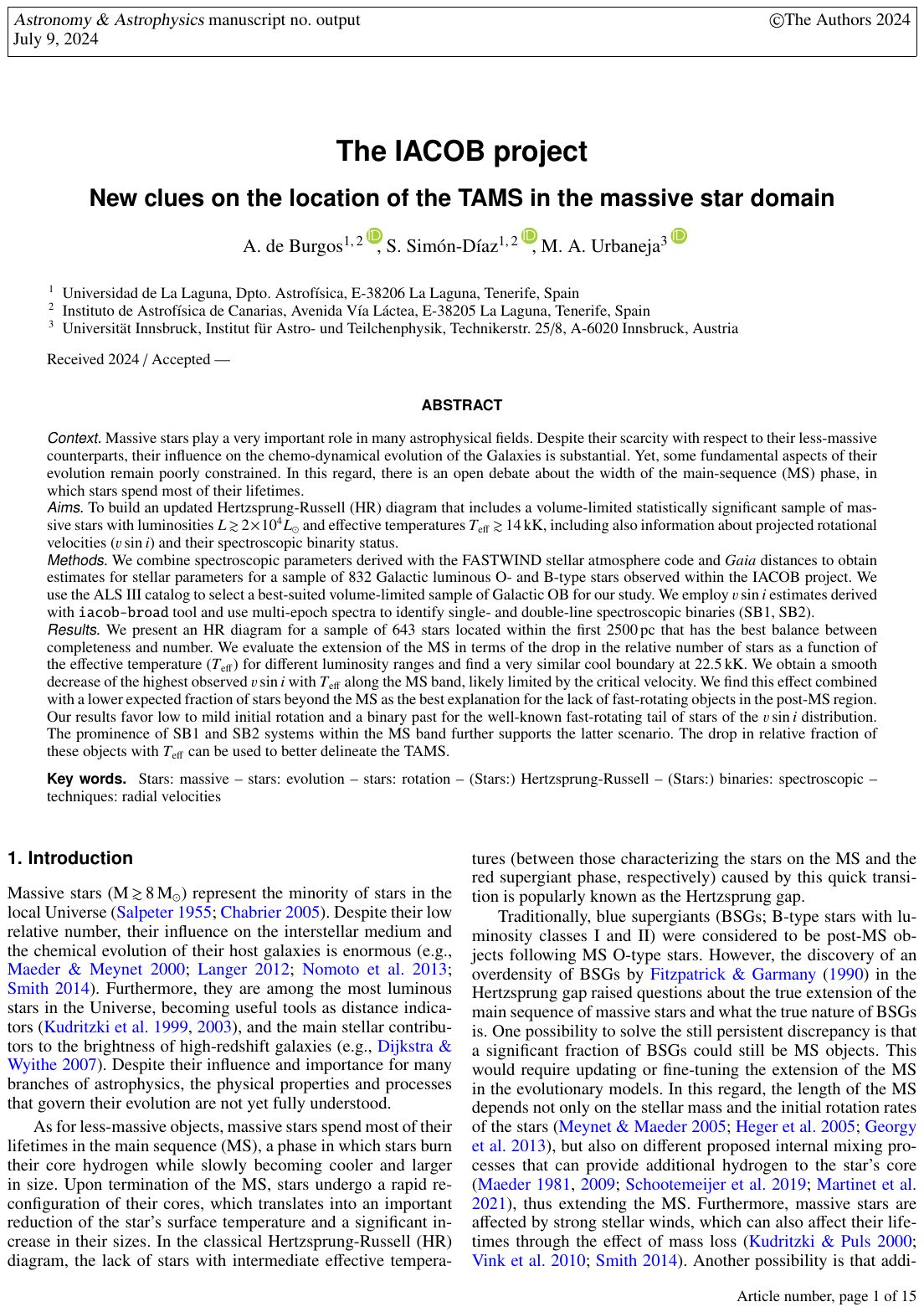}

\clearemptydoublepage{}

%
%
\chapter{On the properties and evolutionary nature of Galactic B-type supergiants around the TAMS}
\label{chapter5}

{\flushright{\it He is glorified not in one, but in countless suns, not in a single earth... \\...but in a thousand thousand, I say in an infinity of worlds.\\\smallskip
            }{\rm \small Giordano Bruno\\}}

\vspace{0.7cm}
%
%
\begin{center}
This chapter includes the content of a paper at an advance stage\\of preparation that will be submitted to Astronomy and Astrophysics.

\end{center}
\includepdf[pages=-]{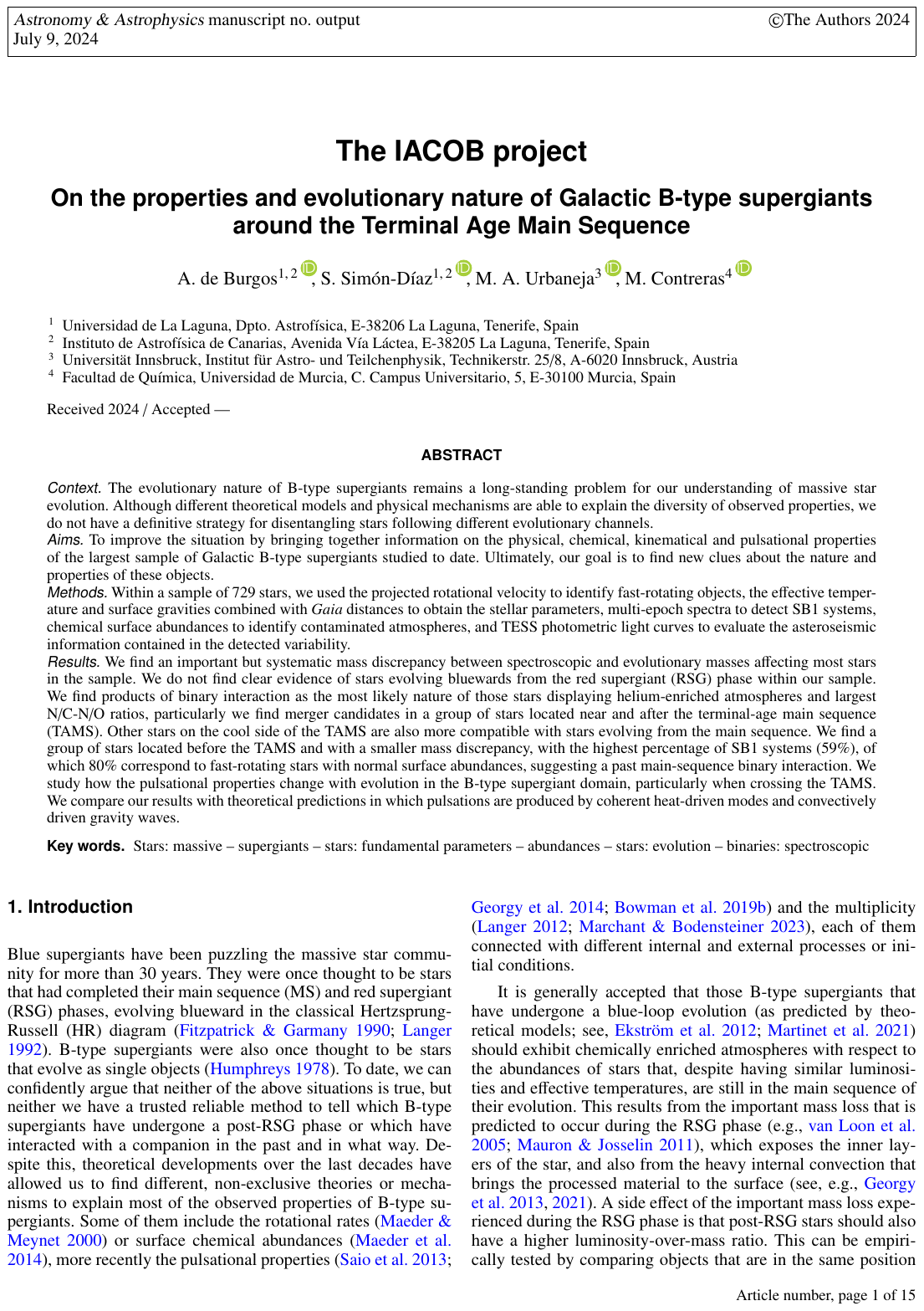}

\clearemptydoublepage{}

%
%
\chapter{Conclusions and future work}
\label{chapter6}
{\flushright{\it Not all those who wander are lost.\\\smallskip}{\rm \small J.R.R. Tolkien, The Fellowship of the Ring\\}}

\vspace{0.7cm}
%
%
\textit{This last chapter summarizes the main results and highlights of the research done in this thesis. It also includes future interesting studies that should follow this work.}\\

\dropping[0pt]{2}{B}{\sc -type} supergiants are very important pieces in the puzzle of massive star evolution theory and in our understanding of astrophysics in general. Despite the theoretical developments of the last decades, discrepancies between observational data and theoretical models have raised long-standing and new questions. Furthermore, B-type supergiants likely include objects of different evolutionary nature, further complicating the interpretation of the results.
To improve this situation, statistically significant and homogeneous samples analyzed with a self-consistent approach are needed.

Despite this fact, until the present work there has not been a study in the Milky Way that has analyzed a spectroscopic dataset as large and of the quality as the one used in this thesis. In fact, at the time of its completion, it comprised $\sim$900 B-type supergiants and more than 6000 high-resolution spectra with high signal-to-noise ratios. 

This important milestone was achieved by conducting several successful and strategic observing campaigns that aimed to improve the completeness of the sample and reduce observational biases. The combination of this excellent dataset gathered within the IACOB project, combined with a wealth of additional information from \textit{Gaia} and TESS space missions, set the core of this thesis, which has provided a modern empirical reappraisal of the properties of B-type supergiants in the Galaxy and added new insights into their evolutionary nature. 

The following pages summarize the achievements and important results obtained in this thesis. They are listed following the same order as they were presented in the different chapters, since, in general, each chapter was built upon the results of the previous one, as would do any scientific study.\\

\vspace{0.3cm}
{\parindent0pt\textbf{\textit{Building a modern empirical database of blue supergiants}}}
\medskip

The article published in Astronomy \& Astrophysics (A\&A) corresponding to Chapter~\ref{chapter1}, was mainly dedicated to present and describe the spectroscopic sample of stars used in the follow-up chapters. We also evaluated its completeness level and provided a general overview on the rotational properties. In particular:

\begin{itemize}
    \item We built a sample of 733 O9- to B9-type stars from an initial set of $\approx$11\,000 high-resolution spectra from $\approx$1600 Galactic late O- and B-type stars. Of them 66\% correspond to blue supergiants and the remaining ones to late O- and early B-type stars of interest. 

    \item We performed the completeness analysis of our sample using the Alma Luminous Star catalog III (ALS~III) and \textit{Gaia}-DR3 data as reference. We obtained a very good completeness level of 80\% for all-sky stars with B$_{\rm mag}$\,$<$9 and 70\% for stars with B$_{\rm mag}$\,$<$11 in the northern hemisphere, both percentages within the first 2\,kpc. Additionally, this allowed us to mine the gaps and reduce biases in the following observing runs.

    \item We developed a novel but simple technique to select stars in a specific region of the spectroscopic Hertzsprung--Russell diagram by fitting the H$\beta$ line, using the widths at different depths as a proxy of gravity, breaking the degeneracy produced by the observed projected rotational velocity (\vsini). This robust method is of much utility with the arrival of the next generation of spectroscopic surveys.

    \item This method also highlighted the risk of relying on spectral classifications from the literature. Many of these classifications, particularly those obtained using low-resolution spectrographs, were found to be incorrect. While we have reclassified the most concerning cases, a thorough reclassification is still urgently needed.

    \item We performed the identification of 51 SB2 and five SB3 systems within the sample by using the available multi-epoch spectra and several key diagnostic lines for this purpose. Of all of them, we discovered eight new systems. 

    \item A morphological classification of the H$\alpha$ and H$\beta$ lines was done for the best spectrum of each star in the sample, revealing first-time showed trends with the spectral types and the luminosity classes.
    
\end{itemize}

\vspace{0.3cm}
{\parindent0pt\textbf{\textit{Large-scale quantitative spectroscopic analysis of the sample}}}
\medskip

A second article accepted in A\&A corresponds to Chapter~\ref{chapter2}, and was dedicated to present the results from the quantitative spectroscopic analysis of the stars selected from Chapter~\ref{chapter1}. We provide an updated overview of the spectroscopic parameters of B-type supergiants, including information on the surface abundances and the stellar winds. The key aspects of the methodology and main findings are as follows:

\begin{itemize}
    \item We used the semi-automatic {\tt iacob-broad} tool to perform the line-broadening analysis of the stars. This tool combines a Fourier transform and goodness-of-fit techniques to provide estimates of \vsini\ and the macroturbulent velocity (\vmac).
    
    \item Supervised machine learning techniques combined with sophisticated atmospheric models (computed with the \textsc{FASTWIND} code) were used to perform a homogeneous spectroscopic analysis of the sample in much shorter timescales compared with traditional grid-based methods. This step also included a rigorous evaluation of the input diagnostic lines and the output probability distribution functions in order to assess and guarantee the quality of the results, which we made public for the community.

    \item The latter technique allowed us to derive effective temperatures (\Teff), surface gravities (\logg), microturbulent velocities (\vmic), surface chemical abundances of helium and silicon, and the wind-strength parameter (\logQ) for a total of 527 O9- to B5-type stars, mostly comprising B-type supergiants. This immediately became the largest sample of Galactic luminous blue stars analyzed to date, surpassing the previous reference work by \citet{castro14}.

    \item We provided a new \Teff\,--\,Spectral-type calibration for O9\,--\,B5 supergiant stars using accurate spectral reclassifications. We show a considerable improvement with respect to previous calibrations that used smaller samples of stars.

    \item Benefiting from the sample size, we performed a first attempt to locate the position of the TAMS based on the drop in number of stars below \Teff\,$\sim$21\,kK. In contrast to the classic interpretation that B-type supergiants are helium-core burning post-main sequence objects, our results suggested that a representative fraction of them would still be on the main sequence.

    \item As for O-type stars \citep[e.g.,][]{conti77, holgado22}, the \vsini\ distribution found for B-type luminous stars has a low-\vsini\ component and a tail of fast rotators (\vsini\ \gs100\kms), which disappears below $\sim$21\,kK. The lack of surface braking of the low-\vsini\ component within the full \Teff\ range, plus the constant percentage of fast-rotating objects and associated upper envelope of values above $\sim$21\,kK, all indicate a very efficient angular momentum transport mechanism between the core and the surface of massive stars. 

    \item Despite the lack of fast-rotating stars and the drop in the \vsini\ distribution below a given effective temperature are well known \citep[e.g.,][]{howarth97, fraser10}, our sample provides stronger empirical evidence of these features with respect to \Teff. The observed drop has been proposed to either delineate the end of the main sequence, or be the result of an enhanced angular momentum loss caused by enhanced mass-loss rates at the predicted bi-stability jump \citep[see][]{vink10}. This motivated us to undertake the work of Chapter~\ref{chapter3}.

    \item We found evidence of the positive connection between microturbulent and macroturbulent velocities, two line-broadening parameters for which the exact physical origin is not well understood. We showed for the first time, a map of microturbulence in a spectroscopic HR diagram \citep[sHR: \Ls\,--\,\Teff;][]{langer14}; and strong correlations between \vmic, \vmac, and \logLs, providing insight into the properties of the associated turbulent motions. 

    \item We found a majority of stars with helium abundances \He\,$\sim$0.10, and a sub-group exhibiting helium-enriched atmospheres (\He\,$>$0.13) representing $\sim$20\% of the sample. After comparing with evolutionary models from \citet{brott11a}, \citet{ekstrom12}, and \citet{keszthelyi22}, we suggested that the origin of this sub-group might be in binary evolution.

    \item The results for silicon abundances provided an average value of $\epsilon_{\rm Si}$\footnote{$\epsilon_{\rm Si} = \texttt{12+}\log\left(\frac{\rm N(Si)}{\rm N(H)}\right)$}\,= 7.46$\pm$0.14\,dex. This is consistent with previous results, but we noticed a larger scatter in our case, which we attributed to the larger sample and important mix of objects.

    \item We showed a clear connection between the wind-strength parameter and the morphological classification of the H$\alpha$ line profile (taken from Chapter~\ref{chapter1}), which we showed for the first time in a sHR diagram. This allowed us to study the region below \Teff\,$\sim$22\,kK, where increased mass-loss rates were predicted \citep{vink01}. Despite we found an important separation of stars with \logQ\ above and below $-$13.6 separating stars with very different H$\alpha$ profiles, we did not find any increase in \logQ.
    
\end{itemize}

\vspace{0.3cm}
{\parindent0pt\textbf{\textit{No increase of mass-loss rates over the bi-stability region}}}
\medskip

The third accepted article in form of A\&A Letter corresponds to Chapter~\ref{chapter3}. The initial aim of this work was to test the mass-loss rate predictions of \citet{vink01} to further explore the possibility that the observed drop in fast-rotating objects described in Chapter~\ref{chapter2} is not the result of a ``bi-stability breaking" \citep{vink10} triggered by enhanced mass-loss rates at $\sim$25\,kK (the bi-stability region), but instead the result of stars reaching the TAMS. In brief:

\begin{itemize}
    \item We combined the results of Chapter~\ref{chapter2} with reliable \textit{Gaia}-DR3 distances and wind terminal velocities from the literature (from UV spectra), to obtain upper limits on the mass-loss rates for 116 Galactic luminous blue supergiants covering 35\,--\,15\,kK in effective temperatures. 

    \item We found no increase in mass-loss rates over the bi-stability region in any of the three luminosity ranges in which we separated our sample. Considering the fact that the predicted increase by \citet{vink01} (by a factor of five to seven) is widely implemented in most evolutionary codes \citep{brott11a, ekstrom12, paxton13}, we suggest an urgent revision of the included mass-loss rate prescriptions \citep[e.g., using the alternatives by][]{krticka21, bjorklund23}.

    \item We investigated the effect of clumping in our results using the latest predictions and results for O-type stars and B-type supergiants using the micro- and macro-clumping formalisms. Using also our own simulations with \textsc{FASTWIND}, we find that the distribution of mass-loss rates (as also of the wind-strength parameter) would, at most, become flatter within the considered range of effective temperatures.
    
\end{itemize}

\vspace{0.3cm}
{\parindent0pt\textbf{\textit{New clues on the location of the TAMS of massive stars}}}
\medskip

The Chapter~\ref{chapter4} of this thesis corresponds to an article about to be submitted to A\&A, which is dedicated to investigate the end of the main sequence in the massive-star domain. The lack of increased mass-loss rates found in Chapter~\ref{chapter3} led us to retake the possibility that the lack of fast-rotating objects can also be used to find the location of the TAMS. Benefiting from an extended sample size, we evaluated the drop with \Teff\ in relative number of stars, fast-rotating objects, and also spectroscopic binaries to delineate the TAMS. Our main results can be summarized as follows:

\begin{itemize}
    \item We extended our sample presented in Chapter~\ref{chapter2} with $\sim$150 additional sources from new observing campaigns, and followed the same methodology to obtain their spectroscopic parameters. In addition, we included 150 O3- to O9-type stars from \citet{holgado18, holgado20, holgado22}. After limiting the sample to objects with low error-over-parallax, our sample comprised $\sim$830 sources between 49\,--\,14\,kK. 

    \item We used the ALS~III catalog introduced in Chapter~\ref{chapter1} to select a sub-sample of stars with the best balance between completeness level and number of sources to evaluate the position of the TAMS. The chosen volume-limited sample comprised 643 stars located within the first 2500\,pc, reaching a \gs60\% completeness for all-sky stars with B$_{\rm mag}$\,$<$11, and becoming most complete and homogeneous spectroscopic sample of Galactic massive stars analyzed to date.

    \item We employed \textit{Gaia} distances to obtain absolute luminosities and show this sample in the HR diagram. The analysis of the density of stars showed a drop below \Teff\,=\,22\,--\,25\,kK for luminosities between \logL\,=\,4.3\,--\,5.7\,dex, which corresponds to initial masses between 12 and 40\MSol, and that we interpret as the location of the TAMS. This is an important improvement over previous studies that attempted to locate the TAMS, particularly above $\sim$25\MSol, where the TAMS was suggested to be at $\sim$10\,kK.

    \item An in-depth evaluation of the factors involved in the scarcity of fast-rotating objects on the cool side of the TAMS led us to conclude that this phenomenon is likely due to a combination of stars reaching near-critical velocity and the result of two compatible scenarios. First, that as stars cross the TAMS, the proportion of fast-rotating stars significantly decreases. Second, if fast-rotating stars are the result of binary interaction during their main sequence, and a spin-down occurs in relatively short timescales, it becomes improbable to observe them after the TAMS.

    \item We observed an increasing number of fast-rotating stars extending beyond the TAMS towards lower \logL\ values. Besides the aforementioned factors, we argue that this could be a consequence of rotationally induced effects that extend the width of the main sequence.

    \item We interpreted the much larger fraction of slowly-rotating stars compared fast-rotating ones (shown also in Chapters~\ref{chapter1} and \ref{chapter2}) and the apparent decrease in \vsini\ of the former group with decreasing \Teff\ as an indicator that massive stars may enter the main sequence with low to mild initial rotations (\vsini/\vcrit\,\ls0.2). This would be further supported under the assumption that fast-rotating stars mainly result from binary interaction, as already suggested by several authors \citep[e.g.,][]{ramirez-agudelo13, demink14, holgado22}.

    \item In addition to the SB2 systems identified in Chapter~\ref{chapter1}, we used all available multi-epoch spectra and, in particular, the peak-to-peak radial velocities, to find all single-line spectroscopic binaries (SB1) within the sample of B-type stars. Combined with the information on O-type spectroscopic binaries from \citet{holgado18}, we evaluated the fraction of these systems as a function of \Teff. Our results showed a significant decrease from 37\% of SB1 systems within the main sequence to 14\% beyond the TAMS, which we consider as an additional empirical clue for our proposed location for the TAMS. We found a similar qualitative behavior for the SB2 systems.
    
    \item Our results provide strong empirical constraints for improving stellar evolution models. Particularly with respect to the width of the main sequence in the massive star domain.
    We suggest a revision of the overshooting parameter for non-rotating models of \citet{brott11a} and \citet{ekstrom12}, and \citet{choi16}.

    \item An important aspect to consider in the interpretation of our previous results comes from the possible ``contamination" by objects that have followed other evolutionary paths than the canonical single evolution. These refer to post-RSGs objects that have undergone a blue-loop evolution \citep[e.g.,][]{stothers75, ekstrom12}, but also stars that have evolved through binary interactions \citep[e.g.,][]{sana12, moe17}. 

\end{itemize}

\vspace{0.3cm}
{\parindent0pt\textbf{\textit{Properties and evolutionary nature of B-type supergiants}}}
\medskip

The scientific content of this thesis is closed with Chapter~\ref{chapter5}, and corresponds to an article in an advanced stage of preparation to be submitted to A\&A. It focuses on studying the properties of B-type supergiants on each side of the TAMS, and finding new clues on their nature for which, on top of the empirical information already presented, we include additional information on the spectroscopic masses, CNO surface abundances, and pulsational properties of the sample. The latter obtained from the TESS photometric light curves. The key points and main results of this work are listed below.

\begin{itemize}
    \item We have employed a sample of $\sim$600 B-type supergiants plus $\sim$130 O9-type classified stars of similar \Teff\ and \logg. As in Chapter~\ref{chapter4}, they were selected to have reliable distances. In total, we accounted for 728 stars.

    \item By calculating the spectroscopic masses and comparing them with the evolutionary ones (\difMs), we aimed at investigating objects with strong indications of having gained or lost mass at some point during their evolution. However, our results first evidenced a systematic discrepancy \citep[known as the ``mass-discrepancy problem"; see][]{groenewegen89, herrero90} going from \difMs\,$\sim$50\% for stars within the main sequence, to $\sim$30\% for stars located after the TAMS.

    \item Assuming stars near the mean discrepancy values on each side of the TAMS as ``normal" and those separated more than 3$\sigma$ as ``peculiar", we evaluated the presence of post-RSGs within our sample, as they are predicted to experience important mass-loss episodes during the RSG phase. Moreover, they are also predicted to exhibit helium-enriched atmospheres. However, our results do not suggest their presence within our sample, having important implications to our understanding of massive star evolution.

    \item We considered helium-enriched and fast-rotating objects as potential post-binary interaction objects. We evaluated their positions in the HR diagram and with respect \difMs\ to identify groups of particular interest based on our current understanding of binary interactions. For this, we also included the information on objects detected as SB1 systems.
    For stars within the main sequence, we found a group of $\sim$30 peculiar stars that includes the largest percentage of SB1 systems in the sample (59\%). Their larger \difMs\ ratio, coinciding with many fast-rotating objects and the normal helium abundances, suggests mass accretion of unprocessed material. Although the nature of their companion is not obvious, one possibility could be a less-massive main-sequence object.
    We also found two separate groups of helium-enriched objects.
    One group located immediately after the TAMS and above the 25\MSol track. On average, the $\sim$20 stars in this group have 10\MSol more than those objects located in the symmetrically opposite position with respect to the TAMS, and are our strongest candidates for being merger products. Multi-epoch observations for 70\% of them also discard the presence of companions.
    Moreover, this test also rules out the possibility that any of these objects is a post-RSG.
    The other group is on the main sequence, but also reaches the TAMS and mixes slowly- and fast-rotating stars. The fast-rotating group comprises at least 43\% of SB1 systems and includes some peculiar stars. Thus, it may comprise mass-gainers from a more evolved star, or mergers.

    \item CNO surface abundances were obtained for a sub-sample of 50 stars following a similar methodology as one described in Chapter~\ref{chapter2}, and including different types of objects to provide a broader overview of the behavior of the abundances across the HR diagram. All stars on the hot side of the TAMS are compatible with stars evolving on the main sequence. At the TAMS and beyond, we find a group of helium-enriched objects with spectroscopic masses larger than that of other stars of the same luminosity. Recent predictions by \citet{menon24} suggest that these stars and some main-sequence fast-rotating objects with high N/O and N/C ratios could only explained by merger products.

    \item We processed and analyzed TESS photometric light curves for 215 stars in the sample. We found similarity between the distribution of the standard deviation of the photometric variability and the predictions for turbulent pressure by \citet{grassitelli15a}. However, this also has similitudes with the map of the wind-strength parameter, complicating the study of the physical origin of the widely detected stochastic low-frequency variability revealed by the periodograms.

    \item The results on the pulsational properties did not reveal significant differences between stars located before and after the TAMS, suggesting that the origin of the stochastic low frequency variability after crossing the TAMS is produced either by the hydrogen-burning shell (since the hydrogen burning at the core is halted), or in the sub-surface convection zone, connected with the macroturbulent motions of the surface.

\end{itemize}

\vspace{0.3cm}
{\parindent0pt\textbf{Future work}}
\medskip

For more than 30 years, the evolutionary nature of B-type supergiants has been a subject of study. Currently, we do now have a trusted and reliable method to determine which of these stars have left the main sequence, undergone a RSG phase, or interacted with a companion during their evolution. However, theoretical advances in the last decades have also proposed several non-exclusive theories or mechanisms to explain many of the observed properties of B-type supergiants.

To overcome this long-standing problem and reconcile the discrepancies between observational data and theoretical predictions, studies using statistically significant and bias-reduced samples of these objects, such as the one used in this thesis, are essential.

Looking ahead, the upcoming arrival of WEAVE-SCIP \citep{dalton20, jin24} and 4MOST-4MIDABLE-LR \citep{chiappini19} large spectroscopic surveys will significantly increase the number of observed B-type supergiants in the Milky Way. The insights gained from this thesis will surely benefit us in the mining and interpretation of these new data, and ultimately improving our understanding of the properties and evolution of massive stars and its intricate puzzle.

Besides the above-mentioned upcoming work, we now list some of the ``open topics" which could not be addressed during this thesis, that will certainly become part of the future work. 

\begin{itemize}
    \item Studying massive stars within stellar clusters or associations offers a valuable approach to improve the present situation, thanks to their shared ages and compositions, which simplifies any comparison. 
    The Perseus OB1 association is an ideal case, as it gathers one of the largest observed populations of massive stars in the Milky Way. To study it, we count with high-resolution spectra of $\sim$200 of the most luminous objects. The blue-to-red SGs ratio can also provide important constraints to evolutionary models \citep[see][]{langer95}.

    \item The large number of Galactic B-type supergiants with erroneous spectral classifications (Chapter~\ref{chapter1}) highlighted the need for a comprehensive reclassification. To address this, we introduced a new grid of B-type standard stars in \citet{negueruela24}. In addition, machine learning techniques could facilitate the classification process compared to traditional ``by-eye" methods.

    \item The flux-weighted gravity-luminosity relationship \citep[e.g.,][]{kudritzki03b} of B-type supergiants is a powerful tool for determining extragalactic distances \citep[e.g.,][]{urbaneja17}, as well as to constraint stellar evolution models \citep{meynet15}. We have not explored this tool within this thesis but we believe that, given the properties of the sample, it will provide new useful insights. 

    \item Our results form Chapter~\ref{chapter3} led the door open to investigate the wind momentum-luminosity relationship \citep{puls96, kudritzki99}. This relationship connects stellar wind properties of massive stars with their intrinsic luminosity, and can be used within our sample to investigate how this relationship changes for sub-groups of stars with different spectral types or luminosity classes.

    \item The simultaneous fitting of optical and UV spectra for our complete sample can provide additional information on mass-loss rates and the wind structure (clumping). Moreover, investigating those rates for stars with effective temperatures below 15\,kK will add new constraints to the newer prescriptions of mass-loss rates by \citet{krticka21} and \citet{bjorklund23}.

    \item In Chapter~\ref{chapter5} we presented results of CNO surface abundances for a sub-sample of 50 stars. Unfortunately, we did not have enough time to extend the analysis to the full sample. However, it becomes clear that this will be one of the key works that will follow this thesis, as it will bring further clues on the nature of the B-type supergiants. Particularly on the presence of products of binary interaction.

    \item The results of Chapter~\ref{chapter5} also evidenced a systematic mass discrepancy observed in most stars within the sample. Whether this discrepancy is related to some of the derived spectroscopic parameters, such as the surface gravity, or it arises from the model computations is something that deserves further investigation. In this regard, improved distances from \textit{Gaia} DR4 will also be valuable to find the source of the discrepancy.

    \item The study of pulsational properties in massive stars offers a unique approach to explore their stellar structure and interior. This thesis presents a preliminary work on some of the properties, but future research will hopefully help us to find new clues on the physical origin of the observed variability.

    \item Last but not least, \textit{Gaia} proper motions, combined with systemic radial velocities, can provide information on the dynamical state of stars. These data can be used to identify runaway objects that have been dynamically ejected from their parental clusters. For instance, the case of stars evolving in binary systems that are kicked by supernova explosion of their more evolved companion.

\end{itemize}

\clearemptydoublepage{}

%
%
\appendix
\chapter{Appendices}
\label{Appendix}
\vspace{0.7cm}


\vspace{-0.1cm}
\section{Developed \textit{pyIACOB} Python~3 package}
\label{apdx.pyIACOB}

This appendix aims to illustrate some of the functionalities of the \textit{pyIACOB} package. Developed within the IACOB project, \textit{pyIACOB} is the first Python package designed specifically for users of the IACOB spectroscopic database. It primarily serves present and future generations of master's and doctoral students who are familiar with the Python programming language. 
The package provides quick control and visualization of IACOB spectra, which have been acquired using the FIES, HERMES, and FEROS high-resolution echelle spectrographs (see Sect.~\ref{intro.3.iacob}). However, some functionalities can also be used with any spectra (synthetic or real), at a wide range of resolutions, loaded as an ASCII file.

\subsection{Database module: Search tool and summary tables}
\label{apdx.pyIB.search}

The {\tt database} module, once linked to the folder containing the IACOB spectra, can be used to search for the best signal- to-noise ratio (S/N), or any spectra above an input S/N value. This module can also be used to create master tables by filtering all the stars in the database by their visual magnitude, coordinates (equatorial or galactic), spectral types, or luminosity classes. It also allows to complement the table with the associated information from \textit{Gaia} DR2 or DR3, given a search radius. Other interesting features include the possibility of querying any object in the SIMBAD astronomical database \citep{weis20}, or to run several sanity checks and fixes on the spectral database.

\subsection{Spectrum module: Visualization and line fitting}
\label{apdx.pyIB.spec}

\begin{figure}
    \centering
    \includegraphics[width=0.99\linewidth]{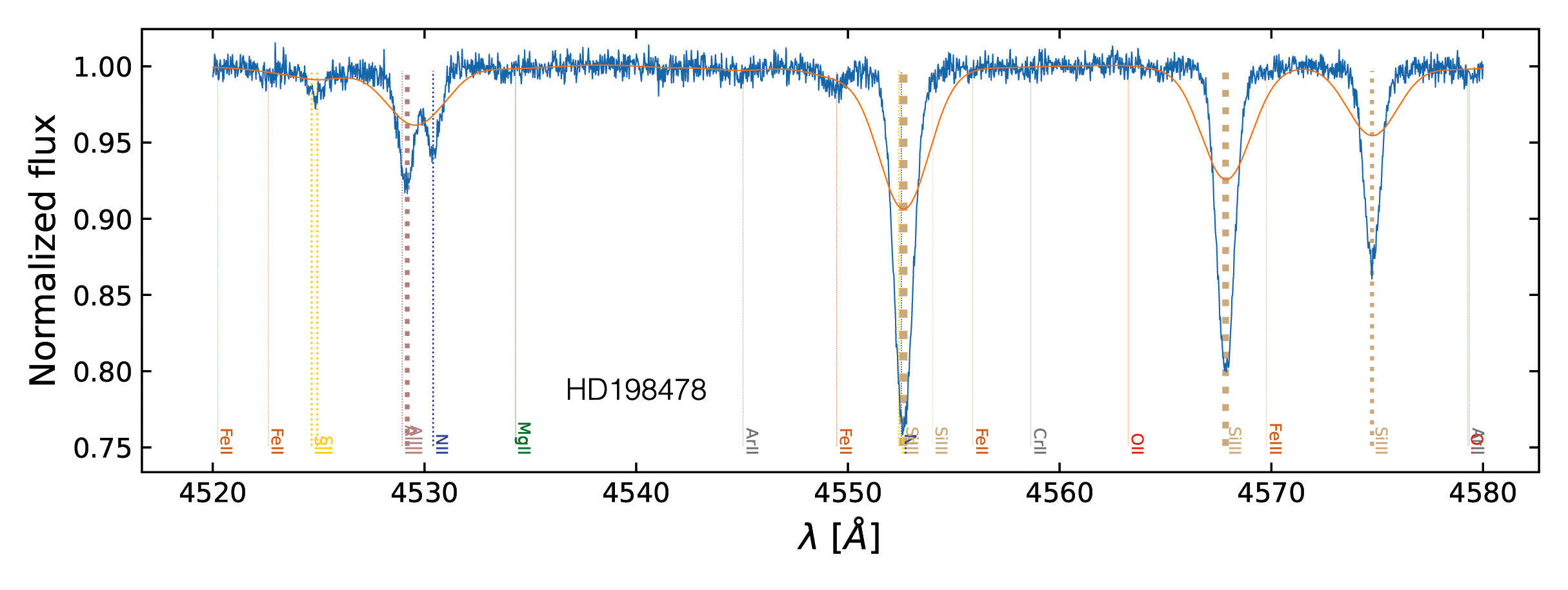}
    \caption{Best S/N spectrum of HD\,198\,478 (B4\,Ia) as shown in \textit{pyIACOB} using the {\tt spec} module. In blue, the original spectrum. In orange, the same spectrum degraded to $R$=2500. Some relevant atomic lines included with the module are shown for reference.}
    \label{fig:apen.plotspec}
\end{figure}

The {\tt spec} module is the most elaborate in \textit{pyIACOB}. It requires from an input spectrum, which can be selected as the best available for a given star. Once done, a class is created with all the relevant information from the file header, and generates the corresponding arrays with the wavelength and the flux. Optionally, the user can provide the associated radial velocity. The module allows to visualize the selected spectrum and overplot different lists of atomic lines (see Fig.~\ref{fig:apen.plotspec} for a sample case). The spectrum can also be cleaned from cosmic rays or other artifacts, degraded to a chosen resolution (see again Fig.~\ref{fig:apen.plotspec}), and exported into an ASCII file.

\begin{figure}
    \centering
    \includegraphics[width=0.99\linewidth]{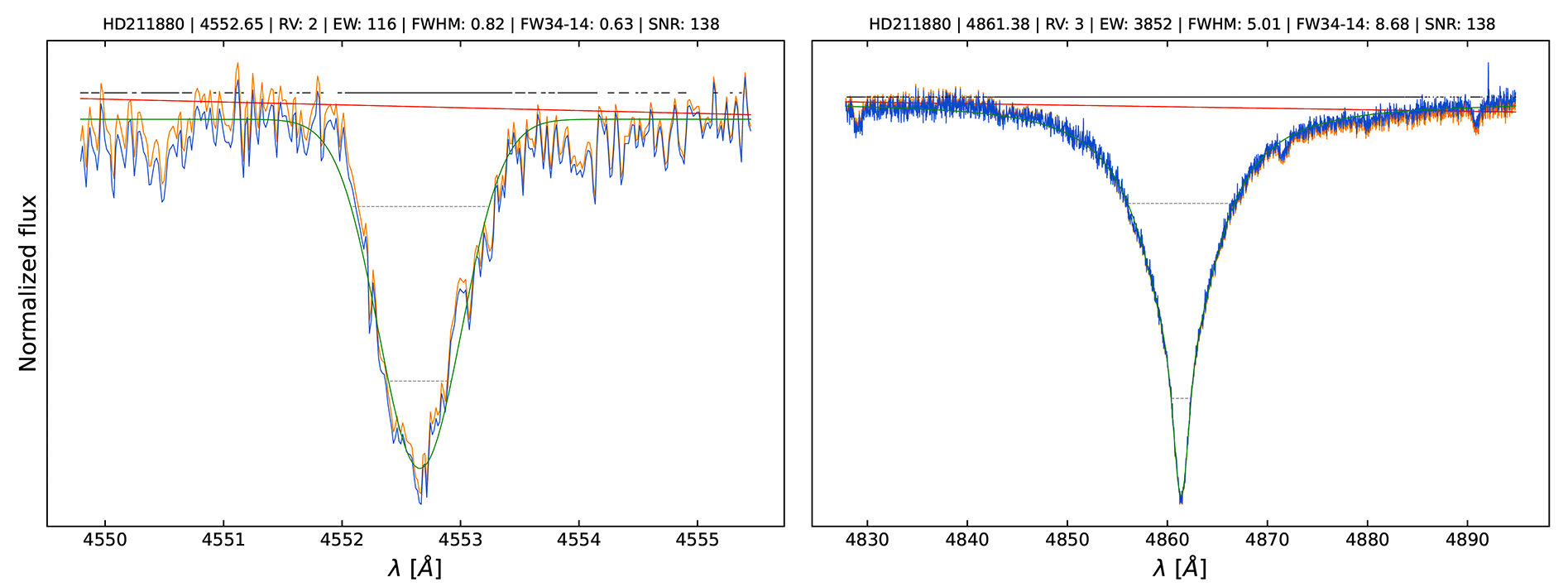}
    \caption{Fitting of the Si~{\sc iii} $\lambda$4552.62\,{\AA} (left) and H$\beta$ $\lambda$4861.33\,{\AA} (right) of HD\,211\,880 (B0.5\,V) as done by the {\tt spec} module of \textit{pyIACOB}. In orange, the original spectrum, and in blue, the normalized one. The green line corresponds to the fitting of the spectral line. The black line on top of the spectrum indicates the areas not used in the normalization. The red line is used for the normalization of the spectrum.}
    \label{fig:apen.fitting}
\end{figure}

In addition to the features mentioned above, the {\tt spec} module allows the user to perform the fitting of spectral lines in absorption using several built-in functions (selected by the user). These include the Gaussian and Lorentzian basic functions, but also other more complex to fit Voigt profiles, and profiles with rotational or macroturbulent broadening (see Fig.~\ref{fig:apen.fitting} for two examples). An iterative local normalization and fitting procedure is also implemented. The module returns the radial velocity of the fitted line, the equivalent width, and the S/N, among other parameters and fitting information.

\subsection{Radial velocity module}
\label{apdx.pyIB.rv}

The radial velocity {\tt RV} module provides \textit{pyIACOB} with the tool to measure the radial velocity of a given spectrum or spectra (as chosen by the user). To achieve this, the user can either provide a list of diagnostic lines or choose one of the default lists optimized for O- B- and A-type stars. A line-fitting routine coupled with a sigma-clipping function is used to obtain the final velocity. Alternatively, a cross-correlation function can be used if a reference spectrum is provided.

\subsection{Higher-level modules}
\label{apdx.pyIB.high-level}

\begin{figure}
    \centering
    \includegraphics[width=0.99\linewidth]{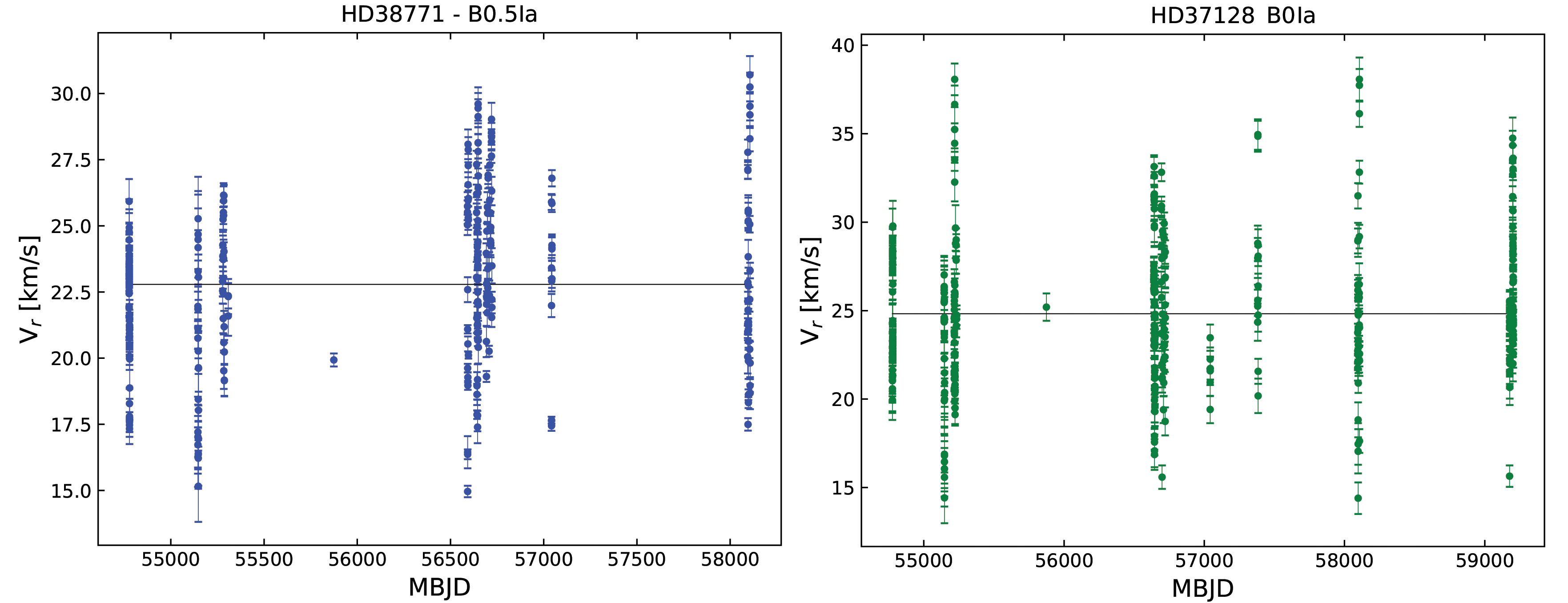}
    \caption{Radial velocity curves of HD\,38771 (left) and HD\,37128 (right) obtained with the {\tt measure.auto\_RV} module of \textit{pyIACOB}.}
    \label{fig:apen.rv}
\end{figure}

A set of modules in \textit{pyIACOB} are based on those already mentioned and are used to facilitate the work when using large samples of stars or spectra. For example, the functions of the {\tt measure} module automate the process of obtaining the equivalent widths of large sets of lines; or produce the radial velocity curves from multi-epoch spectra, which are used to separate intrinsic variability from SB1 systems (see Chapter~\ref{chapter4}). Two examples of radial velocity curves obtained with the {\tt measure.auto\_RV} routine are shown in Fig.~\ref{fig:apen.rv}. The {\tt measure.measure\_Hb} has been particularly useful in measuring the quantity {\tt FW3414\,}(H$\beta$), which is used as a proxy of gravity to separate B-type supergiants (see Chapter~\ref{chapter1} for more details). Additionally, the {\tt binarity} module facilitates the identification of SB2 systems, offering different ways to plot the available spectra. It has been particularly useful in Chapters~\ref{chapter1} and \ref{chapter4}. 

\subsection{Modules connecting Python with IDL programs}
\label{apdx.pyIB.IDL}

Other modules of \textit{pyIACOB} have been created to provide the input or output files of two IDL-based programs used within the IACOB project. The {\tt IACOBroad} module helps the user create the input file for the {\tt IACOB-BROAD} tool \citep{simon-diaz14a}, whereas the module {\tt MAUI} allows to create not only the necessary input files to run the statistical emulator of {\sc FASTWIND}, but also to obtain the probability distribution functions and the best fitting model from the IDL solution files (see Chapter~\ref{chapter2} for more details).

\typeout{}
\bibliographystyle{aa}
\bibliography{biblio.bib}
\addcontentsline{toc}{chapter}{Bibliography}

\clearemptydoublepage{}

\end{document}